\documentclass{aastex62}
\usepackage[figuresright]{rotating}
\graphicspath{{./}{figures/}}

\received{xxx}
\revised{xxx}
\accepted{xxx}
\submitjournal{ApJ}

\shortauthors{Li et al.}

\begin{document}

\title{Exploring the Galactic Anticenter substructure with LAMOST \& Gaia DR2}

\correspondingauthor{Xiang-Xiang Xue; Jing Li}
\email{xuexx@nao.cas.cn; lijing@bao.ac.cn}

\author[0000-0002-4953-1545]{Jing Li}
\affiliation{Physics and Space Science College, China West Normal University, 1 ShiDa Road, Nanchong 637002, P.R.China}
\affiliation{Chinese Academy of Sciences South America Center for Astronomy, National Astronomical Observatories, CAS, Beijing 100012, China}

\author[0000-0002-0642-5689]{Xiang-Xiang Xue}
\affiliation{CAS Key Laboratory of Optical Astronomy, National Astronomical Observatories, Chinese Academy of Sciences, Beijing 100101, China}
\affiliation{School of Astronomy and Space Science, University of Chinese Academy of Sciences, 19A Yuquan Road, Shijingshan District, Beijing 100049, China}
\author[0000-0002-1802-6917]{Chao Liu}
\affiliation{CAS Key Laboratory of Optical Astronomy, National Astronomical Observatories, Chinese Academy of Sciences, Beijing 100101, China}
\affiliation{School of Astronomy and Space Science, University of Chinese Academy of Sciences, 19A Yuquan Road, Shijingshan District, Beijing 100049, China}
\author[0000-0002-6434-7201]{Bo Zhang}
\affiliation{CAS Key Laboratory of Optical Astronomy, National Astronomical Observatories, Chinese Academy of Sciences, Beijing 100101, China}
\affiliation{School of Astronomy and Space Science, University of Chinese Academy of Sciences, 19A Yuquan Road, Shijingshan District, Beijing 100049, China}
\author{Hans-Walter Rix}
\affiliation{Max-Planck-Institute for Astronomy K\"{o}nigstuhl 17, D-69117, Heidelberg, Germany}
\author[0000-0002-3936-9628]{Jeffrey L. Carlin}
\affiliation{AURA/Rubin Observatory, 950 North Cherry Avenue, Tucson, AZ 85719, USA}
\author[0000-0003-1972-0086]{Chengqun Yang}
\affiliation{Shanghai Astronomical Observatory, Chinese Academy of Sciences, 80 Nandan Road, Shanghai 200030, China}
\author{Rene A. Mendez}
\affiliation{Departamento de Astronomia, Universidad de Chile, Casilla 36-D, Correo Central, Santiago, Chile}
\author{Jing Zhong}
\affiliation{Key Laboratory for Research in Galaxies and Cosmology,
Shanghai Astronomical Observatory,
Chinese Academy of  Sciences, 80 Nandan Road, Shanghai 200030, China}
\author[0000-0003-3347-7596]{Hao Tian}
\affiliation{CAS Key Laboratory of Optical Astronomy, National Astronomical Observatories, Chinese Academy of Sciences, Beijing 100101, China}
\author{Lan Zhang}
\affiliation{CAS Key Laboratory of Optical Astronomy, National Astronomical Observatories, Chinese Academy of Sciences, Beijing 100101, China}
\author{Yan Xu}
\affiliation{CAS Key Laboratory of Optical Astronomy, National Astronomical Observatories, Chinese Academy of Sciences, Beijing 100101, China}
\author{Yaqian Wu}
\affiliation{CAS Key Laboratory of Optical Astronomy, National Astronomical Observatories, Chinese Academy of Sciences, Beijing 100101, China}
\author{Gang Zhao}
\affiliation{CAS Key Laboratory of Optical Astronomy, National Astronomical Observatories, Chinese Academy of Sciences, Beijing 100101, China}
\affiliation{School of Astronomy and Space Science, University of Chinese Academy of Sciences, 19A Yuquan Road, Shijingshan District, Beijing 100049, China}
\author{Ruixiang Chang}
\affiliation{Key Laboratory for Research in Galaxies and Cosmology,
Shanghai Astronomical Observatory,
Chinese Academy of  Sciences, 80 Nandan Road, Shanghai 200030, China}


\begin{abstract}
We characterize the kinematic and chemical properties of 589 Galactic Anticenter Substructure Stars (GASS) with K-/M- giants in Integrals-of-Motion space. These stars likely include members of previously identified substructures such as Monoceros, A13, and the Triangulum-Andromeda cloud (TriAnd). 
We show that these stars are on nearly circular orbits on both sides of the Galactic plane. {\textbf{We can see velocity($V_{Z}$) gradient along Y-axis especially for the south GASS members.}} Our GASS members have similar energy and angular momentum distributions to thin disk stars.
Their location in [$\alpha$/M] vs. [M/H] space is more metal poor than typical thin disk stars, with [$\alpha$/M] \textbf{lower} than the thick disk. 
We infer that our GASS members are part of the outer metal-poor disk stars, and the outer-disk extends to 30 kpc. Considering the distance range and $\alpha$-abundance features, GASS could be formed after the thick disk was formed due to the molecular cloud density decreased in the outer disk where the SFR might be less efficient than the inner disk.
 
 
\end{abstract}
 

\section{Introduction} 
{\textbf{\subsection{Historical Overview of Galactic Anticenter substructure stars}}}
The low Galactic latitude substructures collectively known as ``The Monoceros Ring'' (thereafter referred to as Mon) were first discovered by \citet{nyetal02}. Four pieces of stellar over-densities were identified at low latitude -- those labeled S223+20-19.4, S218+22-19.5, and S183+22-19.4 located in the north Galactic cap, and S200-24-19.8 found below the plane. All of these detections are within $40^{\circ}$ of the Galactic anticenter at $(l,b)=(180^{\circ}, 0^{\circ})$.  The turnoffs of the faint main sequence of these overdensities are significantly bluer than the turnoff of typical thick disk stars. \citet{nyetal02} interpreted the bluer turnoff as evidence for a more metal-poor population, as had been identified in the Sagittarius dwarf tidal stream, though the effect could also be due to a population with a younger stellar age.

Within a year, \citet{2003Yanny} and \citet{2003MNRAS.340L..21I} suggested that the low latitude structure was ring-like, and could potentially encircle the entire galaxy (this same structure has been referred to as the Galactic Anticenter Stellar Structure, or GASS, by \citealt{RP2003} and subsequent work). \citet{2003Yanny} traced the structure from $180^{\circ}<l<227^{\circ}$, and noted that it extended 5 kpc above and below the Galactic plane, though the southern portion was 2 kpc further away. They found a velocity dispersion of $\sim 25$ km~s$^{-1}$, which is lower than that of the thick disk (typically 40-50 km~s$^{-1}$), and a scale height that is larger than the thick disk. Based on these observations, \citet{2003Yanny} argued that``the Monoceros ring" is a tidal stream. Simultaneously, \citet{2003MNRAS.340L..21I} pointed out that the ring was not actually traced around the entire Galaxy, but was detected in the range $120^{\circ}<l<200^{\circ}$, on both sides of the Galactic plane. This suggests that the ring could be an artifact in the disk caused by repeated warping, a tidal stream from an accreted satellite, or part of an outer spiral arm. In the same year, \citet{RP2003} showed that the ring was visible in 2MASS M giant stars, establishing that there is a range of metallicities present in this structure, and believed it was the result of a tidally disrupting dwarf galaxy (for more details about Monoceros/GASS, see \citealt{YannyNewberg2016} and other chapters in \citealt{2016N}). 

Several authors have raised the hypothesis of a flare and warp of the outer disk to explain the Monoceros Ring \citep{2006A&A...451..515M,2006MNRAS.368L..77M,2009A&A...493...71C,2011A&A...527A...6H,2002A&A...394..883L,2014Natur.509..342F}. However,  this idea has difficulties explaining the narrow radial velocity dispersion of Mon \citep{2012ApJ...753..116M}.

An extensive photometric and spectroscopic study using SDSS, including SEGUE and SEGUE-2 spectroscopy in the anticenter region, was carried out by \citet{2012ApJ...757..151L}. They conclude that the Monoceros structure has a metallicity of [Fe/H] $\sim -0.80\pm 0.1$,  and the ring has a higher metallicity than the halo, but slightly lower metallicity and a narrower velocity dispersion than the thick disk. All kinematics of Monoceros stars, however, show prograde motion, rotating with the disk, and not far in velocity from the circular motion of stars in a flat rotation curve.

Subsequently, the Pan-STARRS1 survey presented a panoramic picture of the anticenter in stellar density above and below the Milky Way plane. Figure 2 of \citet{2014ApJ...791....9S} clearly shows the ``band-like'' structure stretching from $l=100^{\circ}$ to $230^{\circ}$, and covering a large Galactic latitude range of $-35^{\circ}<b<35^{\circ}$ in some regions.

Detailed studies of chemical abundances in Mon/GASS have been carried out by \citet{2010ApJ...720L...5C} and \citet{2012ApJ...753..116M}. \citet{2010ApJ...720L...5C} derived chemical abundance patterns from high-resolution spectra of 21 M giants. The abundances of the $\alpha$-element titanium, and s-process elements yttrium and lanthanum, for these GASS stars are found to be lower at the same [Fe/H] than those for MW stars, but similar to those of stars in the Sagittarius stream, other dwarf spheroidal galaxies, and the Large Magellanic Cloud. From low resolution ($R\sim1000$) data for hundreds of F/G stars in the anticenter at distances dominated ($65\%$) by objects in Mon/GASS, \citet{2012ApJ...753..116M} found [Fe/H] $\sim -1$, which is intermediate between the halo and the local thick disk. They also found significantly different metallicities for stars at the same Galactocentric radii above and below the plane ([Fe/H]$ = -0.65$ vs. $-0.87$ dex, respectively).

\citet{2015ApJ...801..105X} showed that there is a vertical asymmetry in the disk that is a function of distance from the Galactic center, as measured in the direction of the anticenter. They think the oscillation lines up with the position and density of the Mon and TriAnd structures, but it is only apparent in the north. There are excess star counts in both the north and the south, but they are at different Galactocentric distances. 

\citet{liting2017} presented an analysis of spectroscopic observations of individual stars from ``A13'', which they concluded based on positions, distances, and kinematical properties to be an extension of Mon. \citet{2018ApJ...854...47S} studied the stellar population of Mon, A13, and TriAnd, to assess the relative numbers of RR Lyrae and M giant stars, and found that both structures have very low $f_{RR:MG}$, supporting the scenario in which stars in Mon, A13, and TriAnd formed in the MW disk.

{\textbf{\subsection{Historical overview of MWTD in the context of the Milky Way}}}

{\textbf{The metal-weak thick-disk(MWTD) population has been confirmed existence over the past two decades. The first paper about MWTD was from \citet{1985ApJS...58..463N}, which presented a sample of 309 non-kinematically weak-metal candidates in the solar neighborhood. Follow-up research suggested that the low-metallicity and low-eccentricity stars belong to a population that is intermediate in its motion perpendicular to the Galactic plane between that of the thin disk and that of metal-deficient objects of extreme eccentricity and the velocity dispersion of this group stars is consistent with thick disk\citep{1990AJ....100.1191M,1995ApJs...96...175}.}}

{\textbf{The other observational efforts include high/ medium-resolution spectroscopic abundance determinations have addressed the problem(the fractions of stars at low metallicity were substantially smaller) of the existence of a MWTD component in the Solar Neighborhood of Galaxy \citep{1995ApJS...96..175B,1998AJ....115..168C,1998AJ....116.1724M,2000AJ....119.2866B,2005A&A...433..911A}. Some authors interpreted the origin of the MWTD in terms of the debris of a "shredded satellite"\citep{2002ApJ...574L..39G,2005A&A...433..911A}.
The preponderance of evidence acquired prior to 2009 suggested that a MWTD component exists in the Solar Neighborbood, although these analyses generally did not consider stars with such a large lag as likely candidate disk-like stars. As argued by \citet{2009MNRAS.399..166V}.}}

{\textbf{In the period since 2009, a substantial volume of work has been carried out, making use of large samples of stars with medium-resolution(R$\sim$2000) spectroscopy obtained from a variety of surveys, in particular the Sloan Digital Sky Survey(SDSS;\citet{2000AJ....120.1579Y}), as well as the higher-resolution(R$\sim$7500) data from the Radial Velocity Experiment(RAVE;\citealt{2006AJ....132.1645S}) and other sources. During this period, our appreciation of the complexity of the halo has also increased. \citet{2007Natur.450.1020C},\citet{2010ApJ...712..692C}, and \citet{2012ApJ...746...34B} have presented the case that this system is well-described in terms of an inner-halo and outer-halo population. Additional evidence supporting the existence of (at least) a dual halo around the frequency of carbon-enhanced metal-poor stars, metallicity distribution function(MDF) and so on\citep{2012ApJ...744..195C,2013ApJ...763...65A,2014ApJ...788..180C}. Furthermore, \citet{2009ApJ...694..130M} studied 250 stars to presence of a new component of the local halo, with the axial ratio similar in flattening to the thick disk and populated by stars $-1.5<[Fe/H]<-1.0$, however, not  rotationally supported. \citet{2014ApJ...794...58B} using a new set of very high signal-to-noise(S/N$>$100/1), medium-resolution($R\sim3000$) optical spectra obtained for 302 of the candidate "weak-metal" stars selected by \citet{1973AJ.....78..687B}, this work proved the presence of MWTD population, and found 25$\%$ of the stars with metallicities $-1.8<[Fe/H]<-0.8$ exhibit orbital eccentricites $e<0.4$, yet are clearly separated from members of the inner-halo population with similar metallicities. These works have raised new and interesting questions concerning the nature of the formation and evolution of both the disk and halo systems which need a large spectroscopic data to arrive at a widely accepted view.}}


The second data release of the Gaia mission \citep{gaia2018}, in combination with large sample of spectroscopic surveys, provides the opportunity to search the Galactic halo substructures {\textbf{in a wide view}} in 7D phase space (i.e., 6D positions+velocities, plus metallicities). Xue et al. (in preparation) identified substructures with high reliability in energy versus angular momentum space with K/M-giants selected from the LAMOST spectroscopic survey and Gaia DR2. From these large groups of substructures we found 4 ring-like groups around the Galactic anticenter, with kinematic features similar to the GASS. In this paper, we present the kinematic and chemical features of these groups. The paper is organized as follows. In Section 2 we describe LAMOST K/M giant stars and the Integrals of Motion and Friends-of-Friends algorithm. The kinematic features of GASS stars are described in Section~3. We present the chemical abundance features in Section~4. The discussion and conclusion are presented in Section~5.

\section{DATA and METHOD}\label{sec:data&method}
\subsection{Data}
The data used in this work consist of spectroscopically-identified K- and M-giants from LAMOST Data Release 5 (DR5). The LAMOST Telescope is a 4m Schmidt telescope placed at Xinglong Observing Station. This National Key Scientific facility built by the Chinese Academy of Sciences \citep{Cui2012,Zhao2012,Luo2012,Deng2012} has finished the first stage of its regular survey (LAMOST-I, from 2011-2017; including the pilot survey), and provided 9,027,634 low-resolution (R $\sim 1,800$) optical spectra in its fifth data release, of which 8,183,667 are stellar spectra. 

\textbf{K-giant stars are selected from LAMOST DR5 according to the criteria presented in \citealt{liu2014} ($\mathrm{(4000K<T_{eff}<4600K \& logg<3.5) \| (4600K<T_{eff}<5600K \& logg<4)}$). The distances of K-giants are estimated following the Bayesian method described in \citet{Xue14}, which is suitable for distance estimation of halo K-giants due to the adopted fiducials of globular clusters. The typical distance precision of LAMOST halo K-giants is 13\% \citep{Yang2019a,Bird2019}.}


M-giant stars are from \citet{Zhong2019b}, which used a spectroscopic template matching method plus 2MASS+WISE photometric selection to identify 40,000 M-giants from LAMOST DR5. The contamination of the M-giant sample by 894 carbon stars have also been excluded by cross-matching with the latest LAMOST carbon star catalog \citep{ji2016,lyb2018}. The distances of M-giants were calculated through the $(J-K)_0$ color distance relation derived by \citet{li2016}. 

 {\textbf{Since the distance of K- and M-giants are derived from different calculation method and calibration with distant stars. We re-calibrated the distances of K- and M-giants with Gaia DR2 parallax rather than Gaia distances estimated by \citet{2018AJ....156...58B}. Because \citet{2018AJ....156...58B} claimed that their mean distances to distant giants are underestimated, due to the stars have very large fractional parallax uncertainties, so their distance are prior-dominated, and the prior was dominated by the nearer dwarfs in the model. Only stars with good parallaxes ($\delta \varpi/\varpi <20\%$) and good distances ($\delta d/d <20\%$) are used to do the calibration, which allows us to compare parallax with $1/d$, and minimize the possible bias from inverting. Finally, we used halo stars(the sample which deleted all the identified group members) selected out from Xue 2019(in prepare) as the calibration sample. For K- and M-giants there are $15\%$ and $30\%$ bias respectively( detail see in figure 1 of \citet{Yang2019b} ). Therefore, we increase the distances of K-giants by $(1/0.85-1) \sim18\%$ and decreased the distance of M-giants by $30\%$.}}

The proper motions of the K- and M-giants are obtained by crossing match with Gaia DR2 less than $1\arcsec$. LAMOST pipeline provides the heliocentric radial velocities $hrv$ with the typical error of $7km^{-1}$ . The chemical abundance (metallicity [M/H] and abundance of $\alpha$-element [$\alpha$/M]) of LAMOST K- and M-giants are from \citet{Zhang2019}, which used a machine learning program called Stellar LAbel Machine (SLAM) to transfer APOGEE \citep{Majewski17} stellar labels to LAMOST DR5 spectra. The corresponding cross-validated scatters of [M/H] and [$\alpha$/M] at high SNR$_g$ ($\sim$100) are 0.037 dex and 0.026 dex. {\textbf{All stars for K- and M-giants did the initial cut, only stars at $|Z| > 5$~kpc and those with 2 kpc$< |Z| <$ 5 kpc and [M/H] $< -1$ are classified as halo stars which we will use to identify substructures in the following section. The disk stars we used in this work as comparison with our GASS members are selected from LAMOST K-giants which $|Z|<3$ kpc (detail in section 3.3).}}

\subsection{Identification of Galactic Anti-center Substructure}

Substructure in the Galaxy can be taken as stars moving on similar orbits, but possibly on quite different orbital phases. The orbit can be characterized by its integrals of motion ({\it I.o.M}). Under the assumption that the potential of Galactic halo, in the simplest approximation, is relatively spherical, {\textbf{there are four integrals of the motion: the energy $ \rm E$, the angular momentum vector $\vec{L}$( $\vec{L_x}$, $\vec{L_y}$ and $\vec{L_z}$)}}. As described in \citet{Yang2019b}, Xue et al. in preparation identified Galactic substructure through grouping stars with similar E and $\vec{L}$ from LAMOST K-/M-giants, SEGUE K-giants and SDSS BHBs. Specifically, the E and $\vec{L}$ can be translated to eccentricity $e$, semi-major axis $a$, the direction of the orbital pole $(l,b)_{orbit}$ and the direction of apocenter $l_{apo}$ (i.e. the angle between apocenter and the projection of x-axis on the orbital plane). Please note that $l_{apo}$ changes with periods, but keeps constant within one period, which can be used to distinguish stars in the same stream but involving in our Galaxy in different epochs (e.g. Sgr leading and trailing arms).

By defining the orbit-likelihood-distance to measure how close two stars distribute in $I.o.M.$ space, Xue et al. in preparation applied friends-of-friends (FoF) algorithm to link stars moving on similar orbits together, where the choice of "linking length" is to make sure Sgr streams can be identified as complete as possible.

{\textbf{There are four data set used in Xue et al. work to identify halo substructures, include LAMOST K-giants and M-giants, SEGUE K-giants and SDSS BHBs.}}
Finally, Xue et al. in preparation identified 27 groups from LAMOST K-/M-giants and SDSS K-giants and BHB stars with group members larger than 50. By comparing the sky coverage of these groups, we find four groups from LAMOST K-/M-giants located in Galactic anti-center which share the same kinematic and chemical properties. Two of these four groups are from the K-giants sample, and the others from the M-giants sample. Their location in Galactic coordinates can be seen in Figure~\ref{ps1}. {\textbf{These four groups share similar kinematic and chemical features on the two sides of the Galactic disk (in the following section, we will discuss the kinematic and chemical features in detail). If we relax the restriction of link-length from the FoF method in Xue et al. (in preparation), the two groups of K-giants will merge into one group because they located in two sides of Galactic disk, this condition is same to two groups of M-giants.  Considering these four groups also share same kinematic and chemical properties, we infer that they belong to the same structure.}} By comparing these candidates with results from various works \citep{nyetal02,2003MNRAS.340L..21I,2003Yanny,2012ApJ...757..151L,2014ApJ...791....9S}, we find that our candidates actually include at least the three components Mon, A13, and Triand. Mon is the ring-like substructure detected in the lower latitudes near the Galactic anticenter, with line of sight velocity similar to the thick disk but much smaller velocity dispersion \citep{nyetal02,2003Yanny,2012ApJ...757..151L}. A13 is a substructure found to the north of the Galactic plane in the anticenter direction, with heliocentric distance $\sim 10-20$ kpc, line of sight velocity distribution similar to the disk, and $[Fe/H]\sim -0.4-0.8$ dex \citep{2010ApJ...722..750S,liting2017}. The TriAnd overdensity is found below the Galactic plane near the anticenter direction, with heliocentric distance $\sim 10-25$ kpc, line of sight velocity distribution similar to a disk model with rotation velocity 150~km~s$^{-1}$, and distance $\sim20$~kpc \citep{2014MNRAS.444.3975D}. The truth is that there are many substructures in the Galactic anticenter region, and many of them overlap in their properties (e.g., spatial distribution, kinematics, and chemical abundances). Our work collects them all as one group with K-/M-giants. 

Throughout this work, we refer to the collection of these structures as the Galactic Anticenter Substructure Stars (GASS thereafter), with the caveat that the features may not all be part of the same structure. In the next section, we will exhibit the features of these candidates in detail, including spatial, kinematic and chemical abundance features. The observational parameters and calculated orbit parameters of the 589 GASS members are listed in Table~\ref{t_catalog} and ~\ref{t_orbs} separately. In what follows, we will further explore the nature of GASS using this sample.

\section{THE kinematic features of GASS}\label{sec:Mon}
\subsection{Spatial distribution}
 
\begin{figure}
\includegraphics[width=0.9\columnwidth]{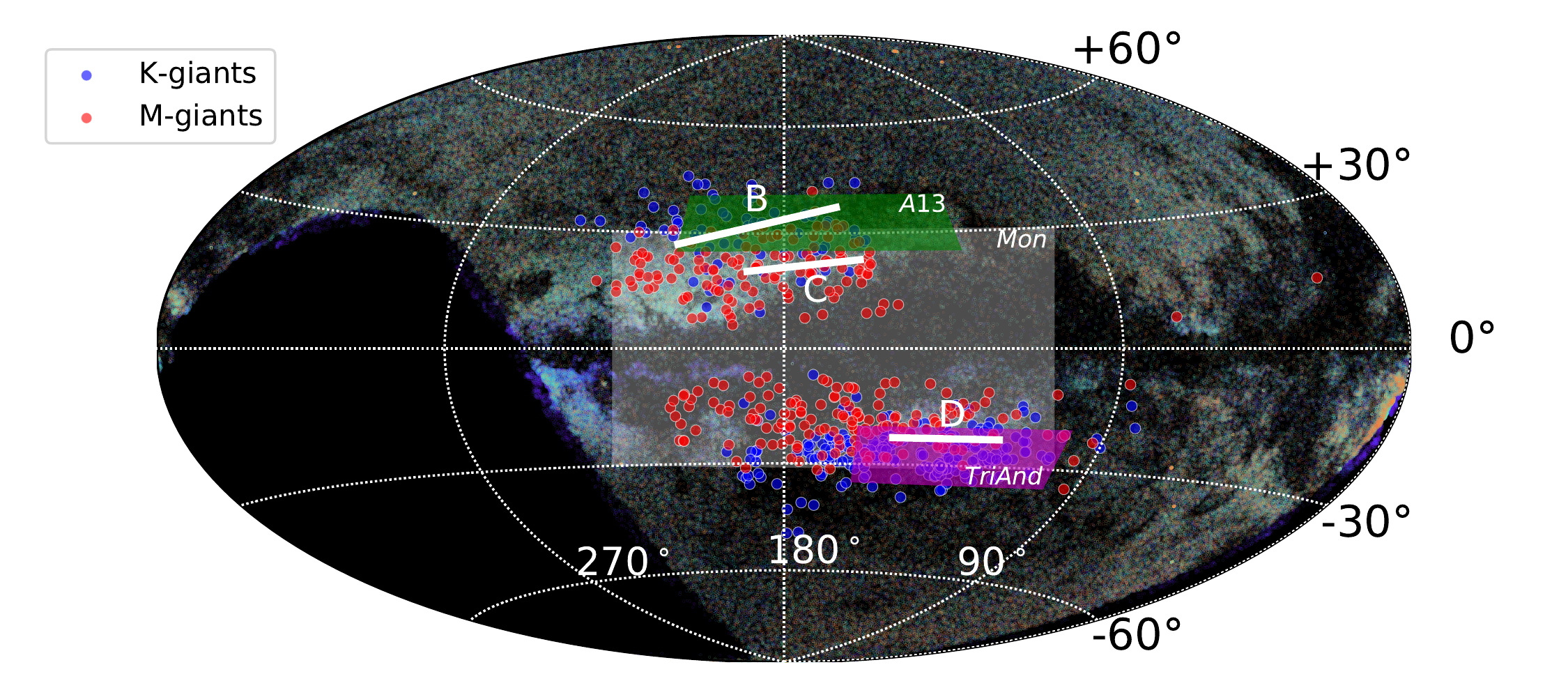}
\caption{{\textbf{Sky coverage of our GASS members in Galactic coordinates. The blue and red filled circles show our GASS Members with K-/M-giants. The background shows sky coverage map of main sequence turn off stars from the Pan-STARRS catalog with $0.2<(g-r)_0<0.3$. Nearby stars with $17.8<g_0<18.4~(4.8-6.3$ kpc) are shown in blue, stars with $18.8<g_0<19.6~(7.6-11.0$ kpc) are shown in green, and more distant stars with $20.2<g_0<20.6~(14.4-17.4$ kpc) are shown in red. The green/white/purple square regions show previous detected A13, Mon and TriAnd regions.}}}
\label{ps1}
\end{figure}

 Figure~\ref{ps1} summarizes the Galactic distribution of the selected GASS members. The members cover a large area on two sides of the disk near the Galactic anticenter, spanning the range $90^{\circ}<l<230^{\circ}$ and $-40^{\circ}<b<40^{\circ}$. We further compare the positions of these stars with the density map of main-sequence turnoff stars from the work of \citet{2014ApJ...791....9S} based on the Pan-STARRS catalog. Figure~\ref{ps1} is similar to Figure~3 from \citet{2014ApJ...791....9S}, but using colored dots that label stars at different distance. Our GASS members have good positional alignment with the "band-like" structures in the Pan-STARRS map. It is worth noting that our sample is in good agreement with the GASS features highlighted in their paper (see, e.g., features B, C, and D in Figure 3 of \citet{2014ApJ...791....9S}). {\textbf{We also labeled out previous detected Mon,A13 and TriAnd regions with white, green and purple squares separately as comparison.}}


\subsection{The Phase-space Distribution}

{For the measurement errors of our sample, LAMOST K-giants have a median distance precision of 13$\%$ \citep{Xue14}, a median radial velocity error of 7 km~s$^{-1}$}, a median error of 0.14 dex in metallicity, and a median [$\alpha$/Fe] error of 0.05 dex \citep{liu2014,Zhang2019}. LAMOST M-giants have a typical distance precision of 20$\%$, but do not have estimates of the distance error for each star \citep{li2016}. LAMOST M-giants have typical radial velocity errors of about 5 km~s$^{-1}$ \citep{Zhong2019b}, a median error of 0.17 dex in metallicity, and a median [$\alpha$/Fe] error of 0.06 dex \citep{Zhang2019}. The proper motions of K-giants and M-giants are derived from $Gaia$ DR2, which is good to 0.2 mas yr$^{-1}$ at G=17$^m$.  In this work, we calculate the errors of each calculated parameter for all K-/M-giants, based on the observational parameters (such as rv, pmra, pmdec, heliocentric distance) using an MCMC method. This involves running our algorithm 1000 times for each single star, sampling from the parameter error distributions (assumed Gaussian) to get per-star errors for the different parameters.
With these parameters and corresponding errors, we are able to analyze the phase-space distribution of the 589 GASS members; the parameters and corresponding errors can be found in Table~\ref{t_catalog} and Table~\ref{t_orbs}.  
\citet{2015ApJ...801..105X} proposed that Mon (close to the Sun) and TriAnd (farther from the Sun) could be associated with the same locally apparent disturbance, as the northern and southern parts of a vertically oscillating ring propagating outward from the Galactic center. \citet{gomez2016} and \citet{laporte2018} using N-body and/or hydro-dynamical simulations have shown that Milky Way satellites could produce strong disturbances and might lead to the formation of vertical structure in the Galactic disk. Some observational work has also shown that there exists a ripple pattern and perturbed velocities in the disk within $r_{gc}<12$~kpc \citep{2017ApJ...835L..18L,2018Natur.561..360A,2018ApJ...865L..19T,2018MNRAS.481.1501B,2019ApJ...872L...1C,2019MNRAS.486.1167B,2019MNRAS.485.3134L}. Figure~8 from \citet{liting2017} schematically illustrates a possible scenario where Mon, A13, and TriAnd are the signatures of disk oscillations at different Galactocentric distances. 

\begin{figure}
\centering
  \begin{tabular}{@{\hspace{-0.5cm}}cccc@{}}
    \includegraphics[width=0.4\textwidth]{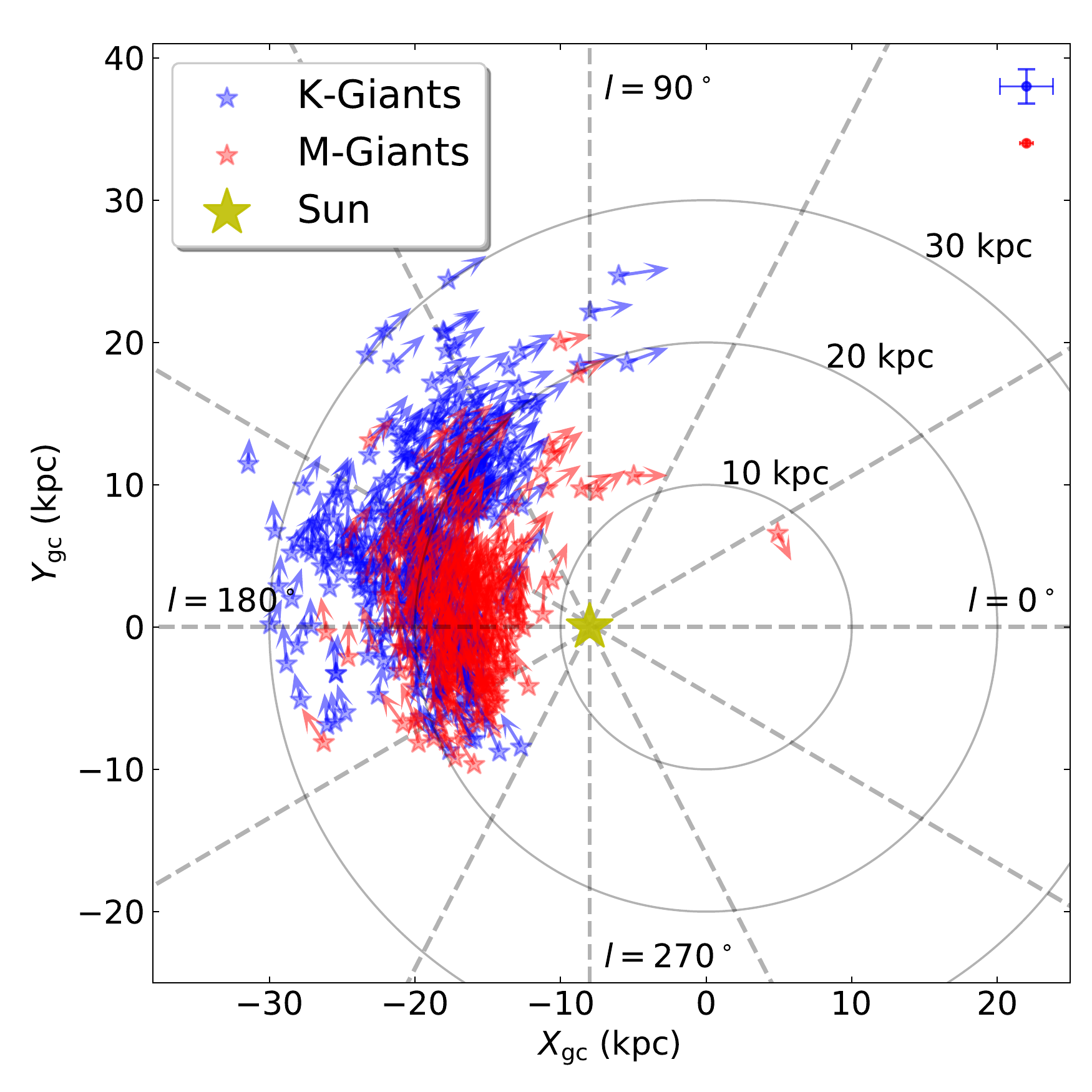} 
    \includegraphics[width=0.45\textwidth]{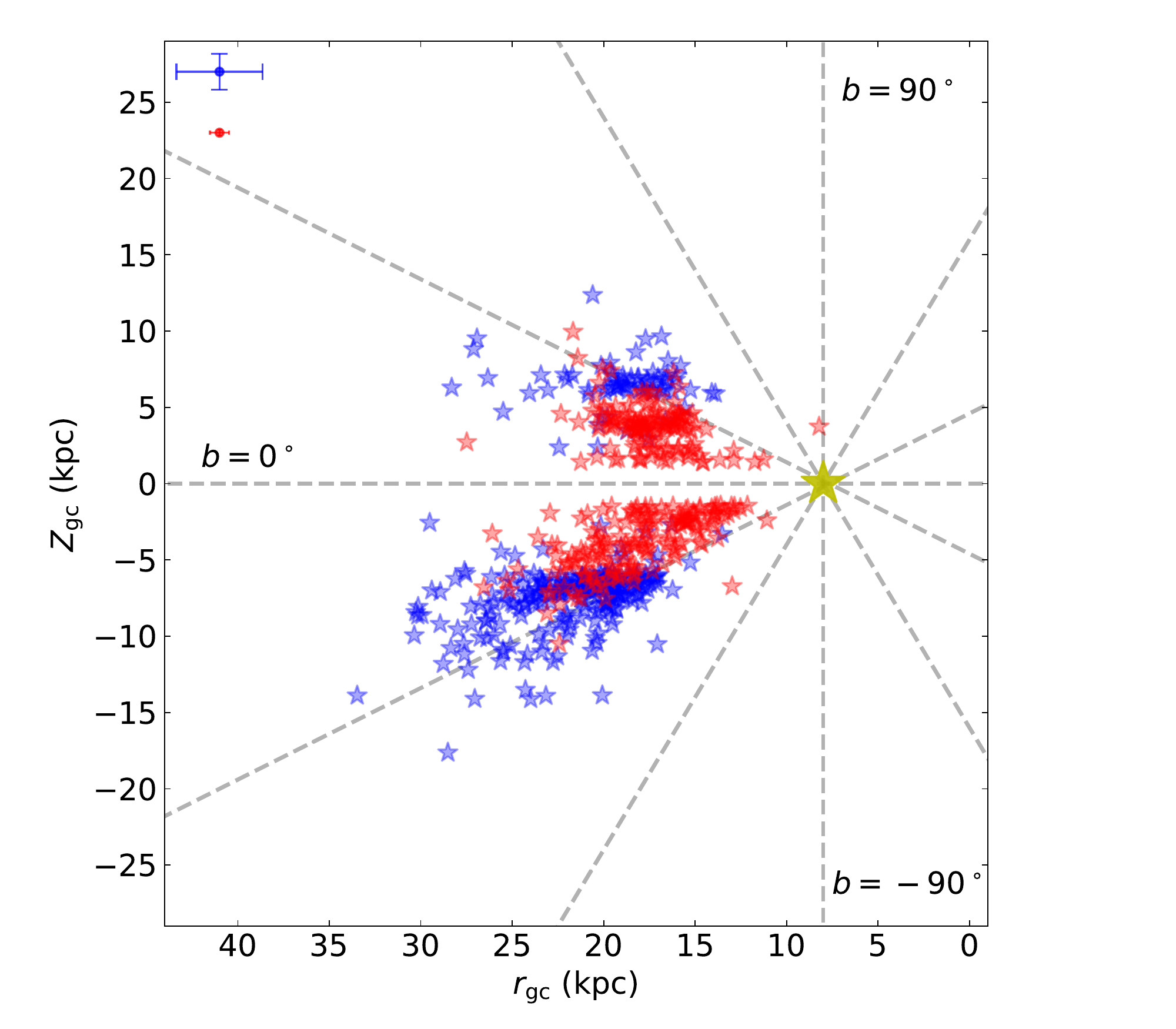} \\
    
  \end{tabular}
\caption{Spatial distribution of the candidates in the X-Y plane (left panel) and R-Z plane (right panel), where r$_{gc}$=$\sqrt{x^2+y^2}$. The Galactic center is at (0,0,0) and the Sun is at (-8.3,0,0) kpc. Galactic longitude and latitude (dashed) and curves at constant Galactocentric radius (solid) are shown.}
\label{xyrz}
\end{figure}

Figure~\ref{xyrz} shows projections of the 3D distribution of our GASS samples onto the Galactic $X-Y$ and $r-Z$ planes{\footnote{The Cartesian reference frame used in this work is centered at the Galactic center, the $X$-axis is positive toward the Galactic center, the $Y$-axis is along the rotation of the disk, and the $Z$-axis points toward the North Galactic Pole. The Sun's position is at (-8.3,0,0) kpc \citep{2016ApJS..227....5D}, the local standard of rest (LSR) velocity is 225 km s$^{-1}$ \citep{2017ApJS..232...22D}, and the solar motion is $(+11.1,+12.24,+7.25)$ km s$^{-1}$ \citep{2010MNRAS.403.1829S}.}}. In the left panel, the arrows indicate the direction and amplitude of velocities in the $X-Y$ plane. {\textbf{We can clearly see our GASS have circle orbit in $X-Y$ plane.}}
In the right panel, we can see that this structure covers a large range in $r_{gc}$, from 15 kpc out to 30 kpc. This range actually overlaps with the distances and part of sky areas to Mon, TriAnd, and A13, since there are not clear boundaries between all of these features. The blue and red error bars in the figure show the mean errors for X, Y, and $r_{gc}$. Because M-giants do not have distance errors for each star, we assign a distance error for M-giants with a random function with restriction of relative error equal 0.05.

{\textbf{Figure~\ref{yz} show spatial distribution of our GASS in the Y-Z plane. In the upper left panel, the arrows represent all GASS members moving direction. We can see a little arc shape from the motion especially in the south hemisphere.}} In the remaining three panels, we color-code regions by the mean $V_{Z}$ component of the stars’ velocities for the samples above/below the plane. From the upper right panel, we see that combined K-/M-giants GASS members show clear {\textbf{$V_{Z}$ gradient along the Y axis}}. The systematical distance error in this work can be ignored because we did calibration for K-giants and M-giants with Gaia parallax. We also make similar plots for K-giants and M-giants separately in the lower two panels. From these two sub-samples we can also see a clear $V_{Z}$ gradient in the Y direction, especially in the M-giants sample, confirming our finding from the combined samples. This behavior {\textbf{could be part of}} expected ripple pattern extending out to $r_{gc}>15$~kpc by \citet{2015ApJ...801..105X} and \citet{liting2017}.

{\textbf{Figure~\ref{vlos} shows the line-of-sight velocity distribution for K- and M-giants members of GASS in north and south hemisphere separately. In the upper panel, shows the north part K- and M-giants groups line-of-sight velocity distribution, the mean velocities for K- and M-giants groups are -20.33$kms^-1$ and -26.87$kms^-1$, the velocity dispersion are 44.08$kms^-1$ and 32.32$kms^-1$. In the lower panel, shows the south part K- and M-giants groups line-of-sight velocity distribution, the mean velocities for K- and M-giants are 17.91$kms^-1$ and 16.27$kms^-1$, the velocity dispersion are 17.31$kms^-1$ and 28.86$kms^-1$. The typical line-of-sight velocity dispersion for thick disk is around 30-40$kms^-1$\citep{2003A&A...410..527B,2012ApJ...757..151L,2014A&A...562A..71B}. The velocity dispersion for thin disk is smaller than 20$kms^-1$\citep{2003A&A...410..527B,2014A&A...562A..71B}. It is hard to simply compare our results to thin and thick disk, there are obviously different velocity dispersion for the north and south parts of our GASS, but entirely the velocity dispersion of north groups are much larger than south groups. This asymmetry structure could be related to the disk ripple/wave structure as expected or detected by previous work.\citep{2014AAS...22334613C,2015ApJ...801..105X,liting2017,2018MNRAS.478.3367W,2020MNRAS.491.2104W} }}

\subsection{Dynamical properties comparison with disk population}

{\textbf{Figure~\ref{LE} shows E vs $L_{z}$ distribution for GASS members(red stars), disk stars(yellow and purple density background) choose from our K-giants sample which $|Z|<3$kpc, Sagittarius stream members selected from \citet{2019ApJ...886..154Y}(green circles). From the density background, we can clear see two components, yellow region are attributed by thin disk stars, the purple region are attributed by thick disk stars.\footnote{The thin and thick disk stars can naturally separate in [M/H] vs [$\alpha$/M] figure as show in figure~\ref{alpha}, we selected the relative pure thin and thick disk stars within 1$\sigma$ distribution for the density distribution for distinct thick and thin disk clump in figure~\ref{alpha}, and check where these pure thin and thick disk stars distribution in E vs $L_{z}$ figure.} As we can see the E vs $L_{z}$ distribution of our GASS members are totally different from thick disk and Sagittarius stream. It is located in an extended narrow region of thin disk stars but have higher E and less $L_{z}$ value than most thin disk population.} }

\section{The Chemical abundance features of GASS}

\subsection{The metallicity distribution}

{\textbf{The [M/H] for all K-/M-giants members spans a large distribution from $-1.50$ to $0.25$, with the mean metallicity around -0.56 dex, and metallicity dispersion around 0.22 dex as shown in Figure~\ref{mh}. From the upper panel, we see that the total K- and M-giants have similar [M/H] distributions.
The lower panel shows that the north and south structures distributions separately, which also have similar [M/H] distributions. For both K- and M-giants, $90\%$ of members the [M/H] value larger than $-1$. Earlier works about Mon/or other Anticenter structures were suggested they could be the remnants of dwarf galaxies which merged in with the outer disk, but for the most Milky Way satellites or dwarf galaxies, the mean stellar metallicites much smaller than -1.5 dex\citep{2012AJ....144....4M,2019ARA&A..57..375S}. Considering the metallicity and E vs $L_{z}$ distribution of our GASS are far from the dwarf galaxies' distribution, So we infer it is unlikely the remnants of dwarf galaxies merged in the MW outer disk.}}

This [M/H] distribution is similar to that found by  \citet{2010ApJ...720L...5C} with high-resolution spectra of 21 M-giants stars. This confirms that the groups in the north and south likely belong to one larger group. 


\subsection{The alpha-abundances distribution}
Figure~\ref{alpha} presents the $\alpha$-element abundances [$\alpha$/M] for LAMOST K- and M-giants obtained by SLAM \citep{Zhang2019}. To compare with the Galactic disk and halo stars, we choose disk stars from LAMOST K-giants with $|Z|<3$ kpc (blue density map in Figure~\ref{alpha}, which naturally separates into thin and thick disk stars in [$\alpha$/M] vs. [M/H] space), and for halo stars we select $|Z|>5$~kpc and eliminate the substructures (blue dots; Xue in preparation). 
As we can see our GASS members have similar $\alpha$-element abundances [$\alpha$/M] as the thin disk, but are more metal-poor than typical thin disk stars. {\textbf{This result is consistent with the continuation of metal-rich thin disk stars into the outer halo \citep{2016A&A...589A..66H}. It is also consistent with the $\alpha$-abundance derived in \citet{2018ApJ...859L...8H} for the TriAnd substructure, which suggests that this feature is an "extension" of the trend seen in the disk.}}


We infer that our GASS members may be part of the outer disk, representing a transition population between the thin disk, thick disk, and halo. 
Figure~\ref{compare_sample} shows K-giant stars selected in the same Galactic distance range as GASS, separated into two samples with $15<r_{GC}<25$ kpc and a more nearby ``outer disk'' sample between $10 < r_{GC} < 15$ kpc. These data are taken from LAMOST K-giants catalogue, but removed all the groups identified in Xue et al. (in preparation) to get a pure sample from the disk and halo. We split these stars into three groups according to $7<|Z|<12$ kpc, $2 <|Z|< 7$ kpc, and a disk sample at $-2 <Z< 2$ kpc. In the upper panel, in the distance range from 10 kpc to 15 kpc, we can see a clear clump of thin disk stars with $|Z|<2$~kpc. From $2<|Z|<7$~kpc, there are still some thin disk stars, but we can also see the clump of the thick disk star sequence. {\textbf{The literature \citet{2012ApJ...752...51C} has suggested that the scale length of the thick disk is quite short, not much more than 2 kpc, whereas traditionally a 3 kpc scale length was assumed. Our result seem to agree with that, duiring the range $10 < r_{GC} < 15$ kpc,$|Z|<2$ kpc are not related to the thick disk at all.}} In the lower panel, showing a distance range from 15 kpc to 25 kpc, we can see that the thin disk stars ($|Z|<2$~kpc) are clearly reduced compared to the more nearby sample. For stars $2<|Z|<7$~kpc, the thick disk star sequence has disappeared (yellow dots [$\alpha$/M]$>$0.2 clump in the upper panel), but the lower sequence still exists, which is similar to the locus of our GASS members in the [M/H]-[$\alpha$/M] space. 

We infer this sequence could be the transition sequence between the thin disk-thick disk-halo. Comparing these two distance ranges, we can see a clear variation in [M/H]-[$\alpha$/M] space within $2<|Z|<7$~kpc. We also see that most thick disk stars appear in the 10~kpc$<R_{GC}<$15~kpc range with $2<|Z|<7$~kpc.

In previous work, \citet{2017ApJ...835L..18L,2018MNRAS.478.3367W} has shown through stellar density profiles that the Galactic disk is more extended than previously thought, reaching out to R$\sim$19 kpc. In Figure 3, \citet{2016A&A...589A..66H} shows there is an inner-disk composed of thick disk and metal-rich thin-disk stars. The $\alpha$ abundance of 12 TriAnd substructure member stars derived in \citet{2018ApJ...859L...8H} shows a similar distribution as our GASS in [$\alpha$/M] vs. [M/H] space; we note that Hayes et al. also claimed TriAnd is an ``extension'' of the disk. 

{\textbf{We also compare our GASS with MWTD, the eccentricites e of MWTD are smaller than 0.4, all our GASS $e<0.35$; the metallicity of MWTD is similar to our GASS, previous detected MWTD are all in the solar neighborhood, but our GASS much further away. Anyway, except the distance range,  the chemical abundance, kinematic feature and eccentricity of our GASS are all very similar to MWTD.}}

Based on the evidence we have presented, we infer that the outer-disk sequence represents a different evolution, where the outer Milky Way still has more cold gas in present day and then maybe lower star formation efficiency than the inner disk. Our GASS members are could be part of the outer-metal-poor-disk stars, and the outer-disk could extend to 30 kpc. 

\section{Discussion and Conclusions}

By combining IoM and FoF algorithms, Xue et al. (in preparation) selected 589 GASS K- and M-giant stars from LAMOST DR5 in the Galactic anticenter region based on their similar kinematic and chemical abundance features. These stars cover a large range in $r_{gc}$, but have similar angular momentum and energy distributions, which could related to the previously identified substructures Mon, TriAnd, and A13, which we have collectively named Galactic Anticenter Substructure (GASS). 

Based on this sample, we present the observations including kinematic and chemical parameters of these stars in Table~\ref{t_catalog} and the calculated orbit parameters in Table~\ref{t_orbs}. All the members are published in on-line readable catalogs.

The GASS covers a large area of the sky, centered around the Galactic anticenter region on both sides of the MW disk, in a Galactic longitude range from from $80^{\circ}$ to $230^{\circ}$, while the Galactic latitude goes from $-35^{\circ}$ to $40^{\circ}$. This coverage is in good positional alignment with the ``band-like'' structures detected in the Pan-STARRS map, especially the high lighted features B,C and D in Figure 3 of \citet{2014ApJ...791....9S}. 
 
\begin{figure}
\centering
    \includegraphics[width=.4\textwidth]{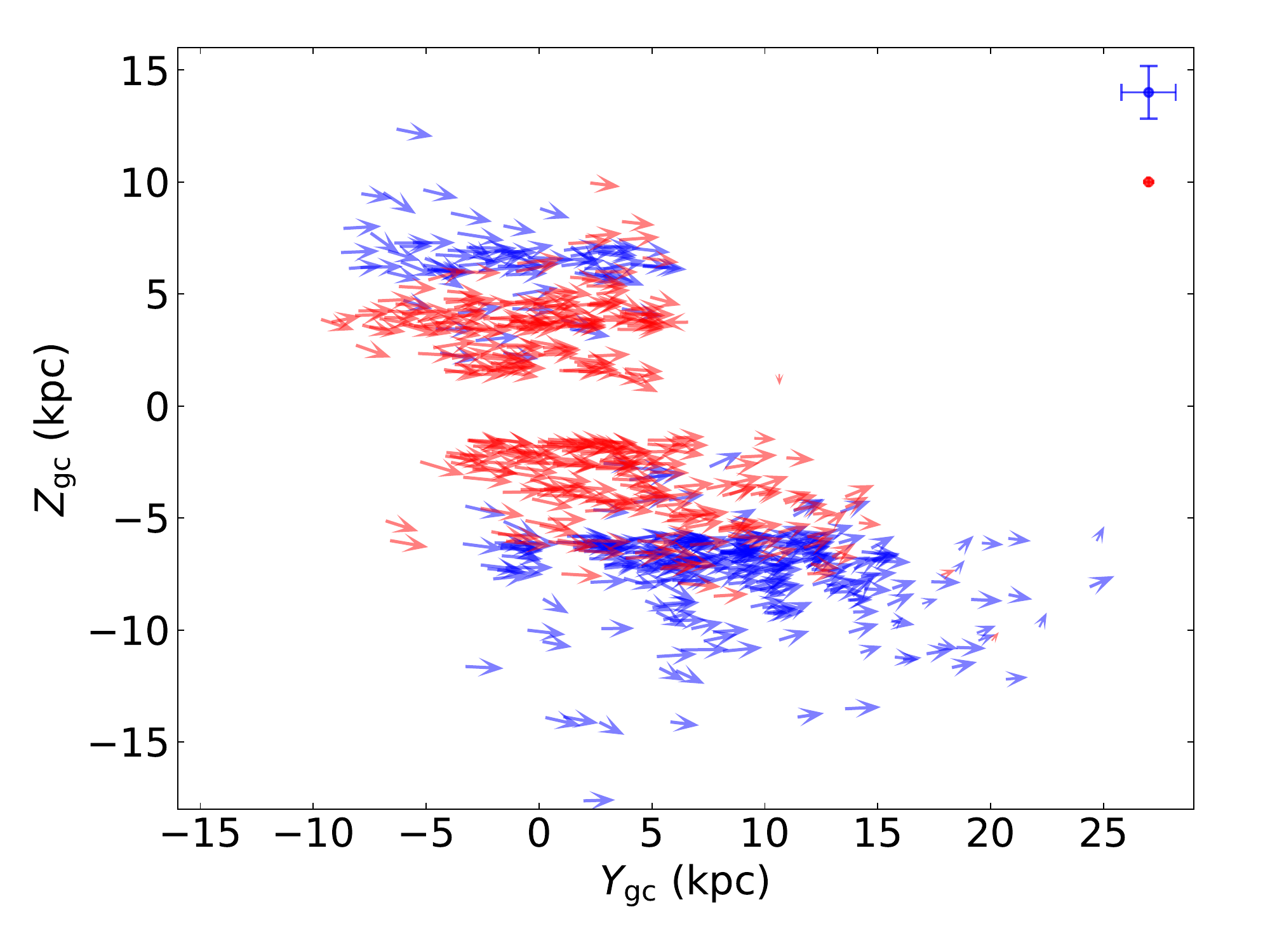} 
    \includegraphics[width=.4\textwidth]{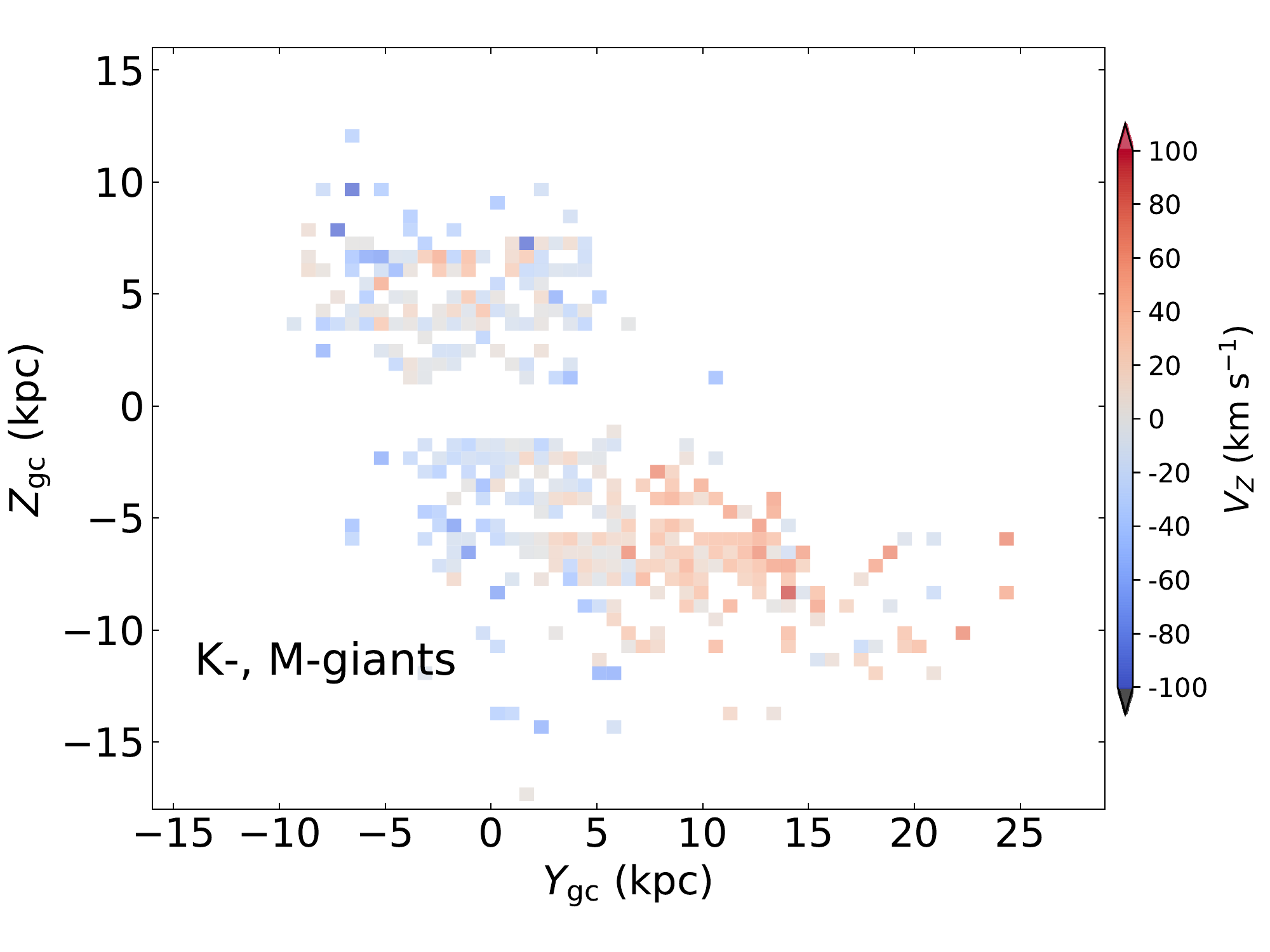} 
    \includegraphics[width=.4\textwidth]{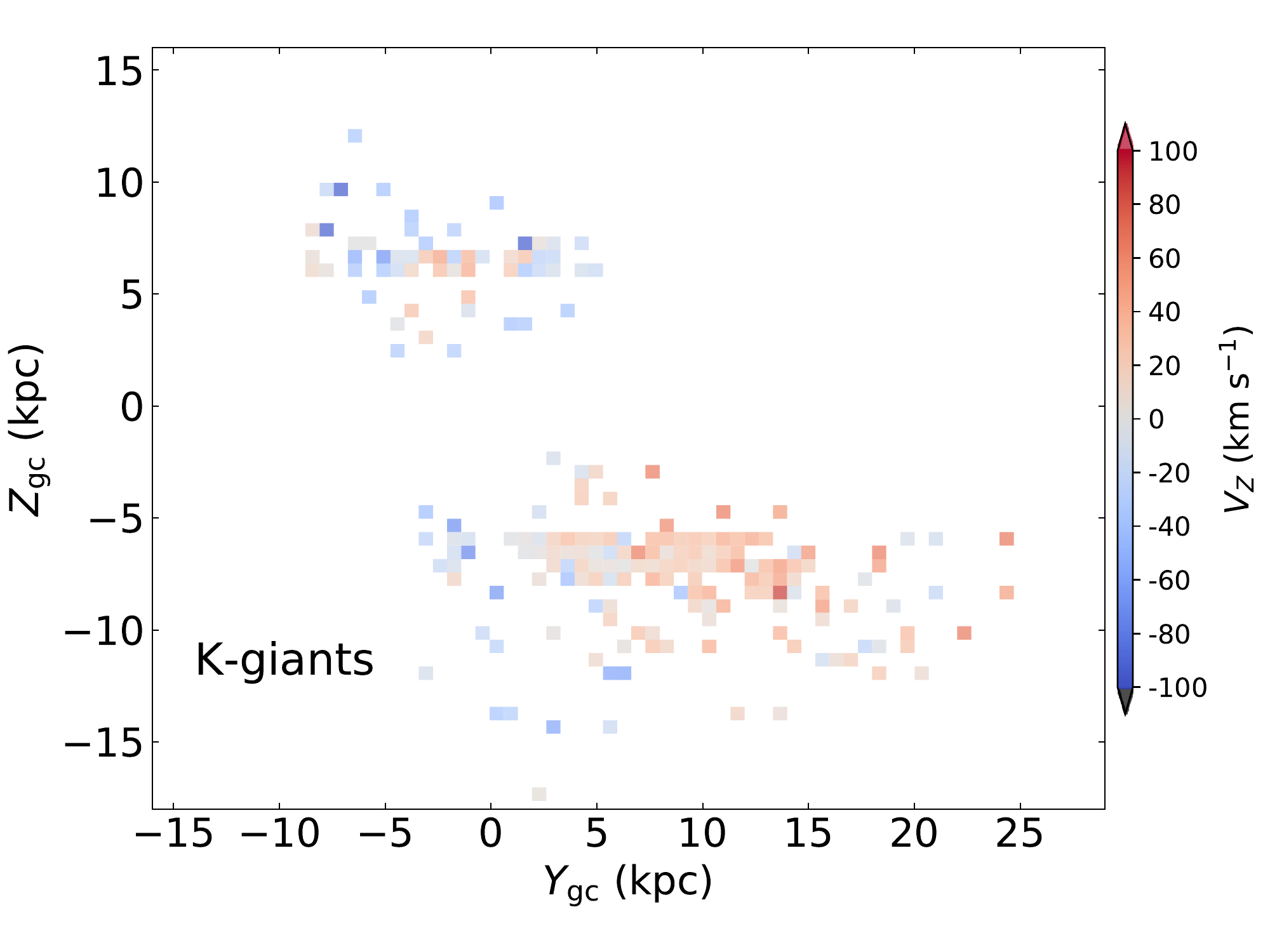} 
    \includegraphics[width=.4\textwidth]{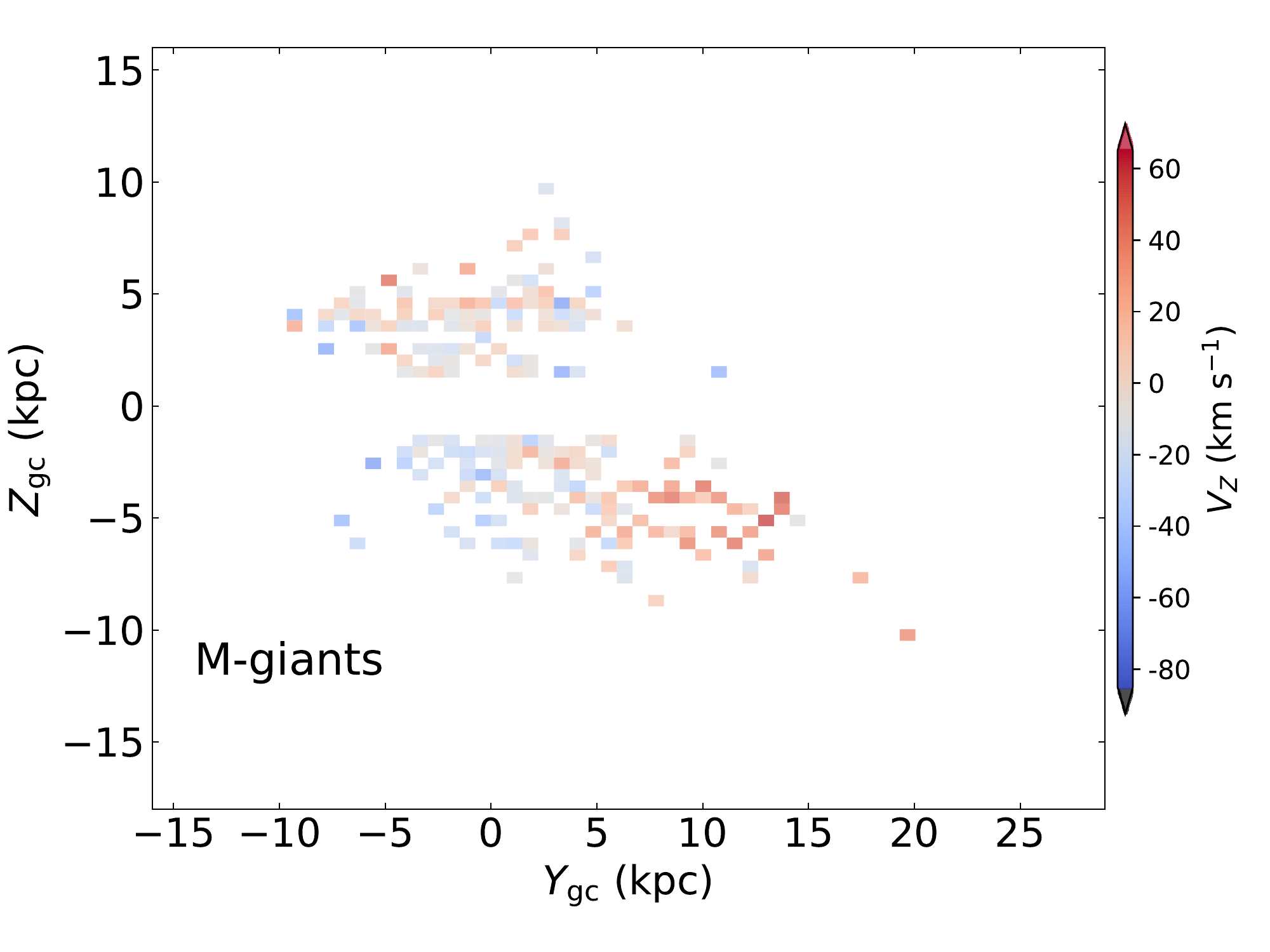} 
\caption{Spatial distribution of the candidates in the Y-Z plane. In the upper left panel, the arrows represent the star´s instantaneous direction of motion and the velocity amplitudes. In the other panels the color shows the mean $V_{Z}$ in each bin in Y-Z space. We can clearly see $V_{Z}$ gradient along the Y axis. Considering that there could be systematic distance errors between K- and M-giants which might affect the oscillation, we illustrate the K- and M-giants mean $V_{Z}$ density maps separately in the lower panels, where the effect is still present.}
\label{yz}
\end{figure}

\begin{figure}
\centering
\includegraphics[width=.5\textwidth]{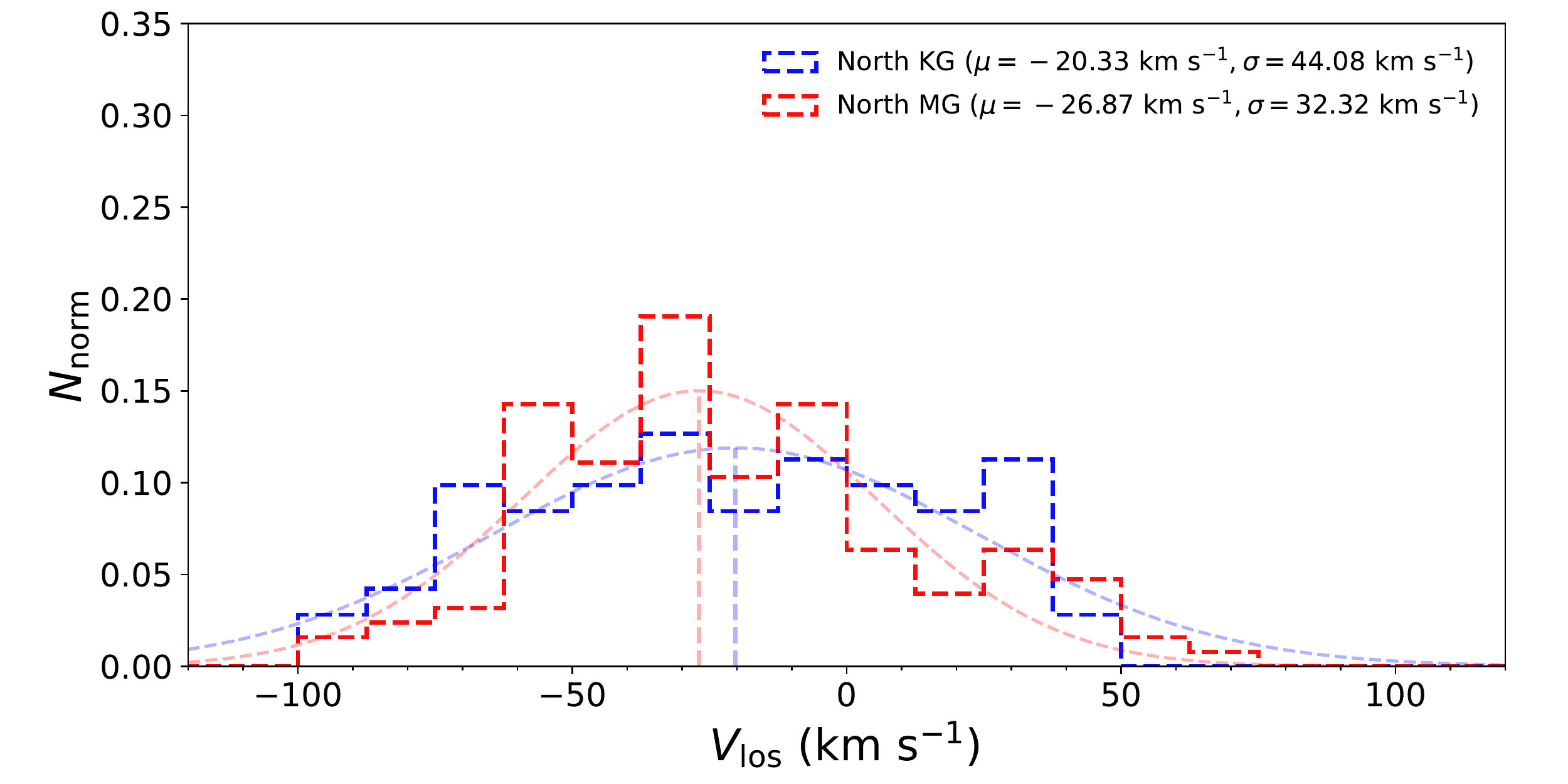} 
\includegraphics[width=.5\textwidth]{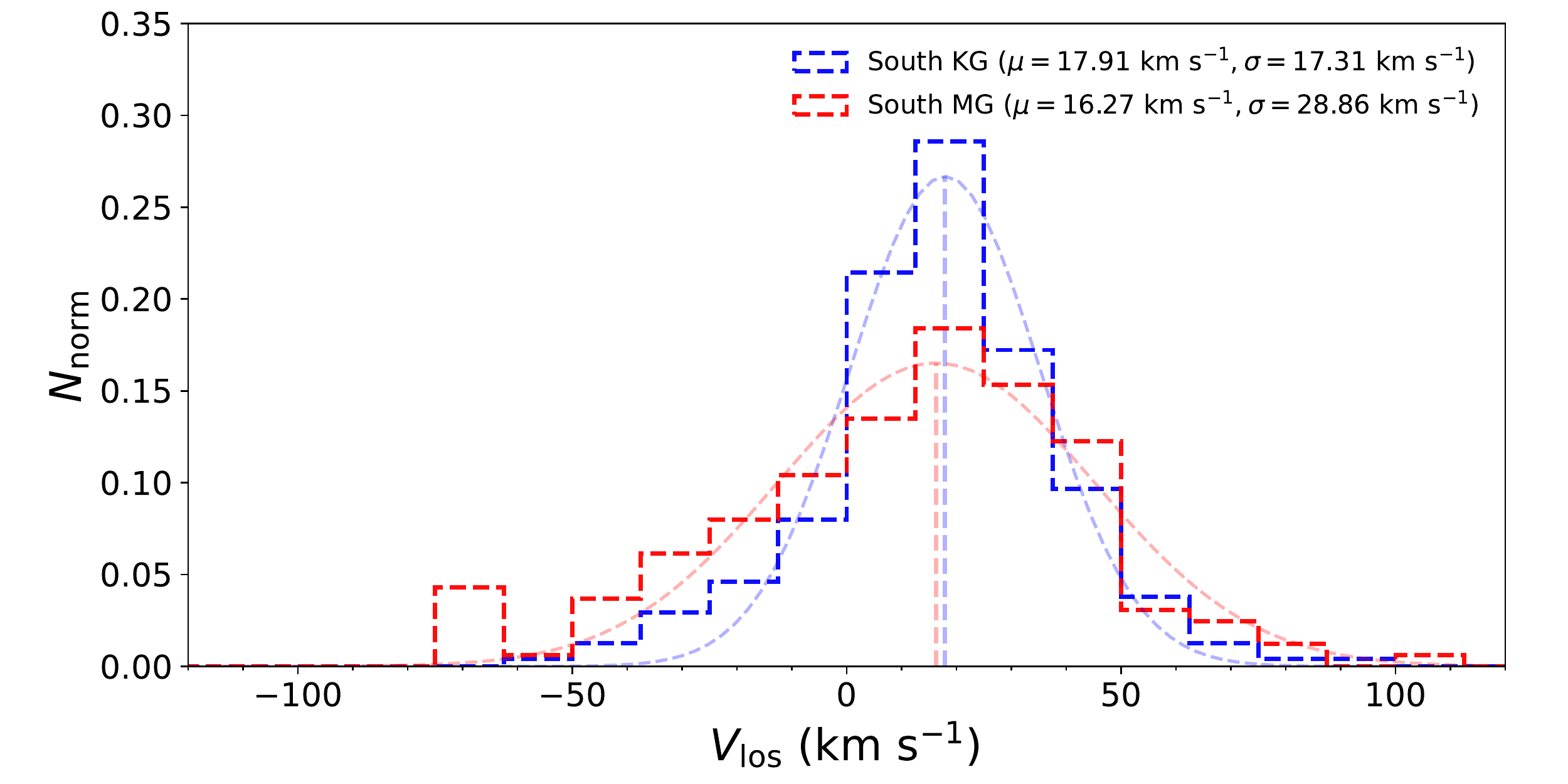} 
\caption{ Line-of-sight velocity distribution for K- and M-giants members of GASS in north and south hemisphere separately. The red and blue dash lines show the Gaussian distribution for each groups, the mean velocities and velocity dispersion show in the figure.}
\label{vlos}
\end{figure}

\begin{figure}
\includegraphics[width=0.8\columnwidth]{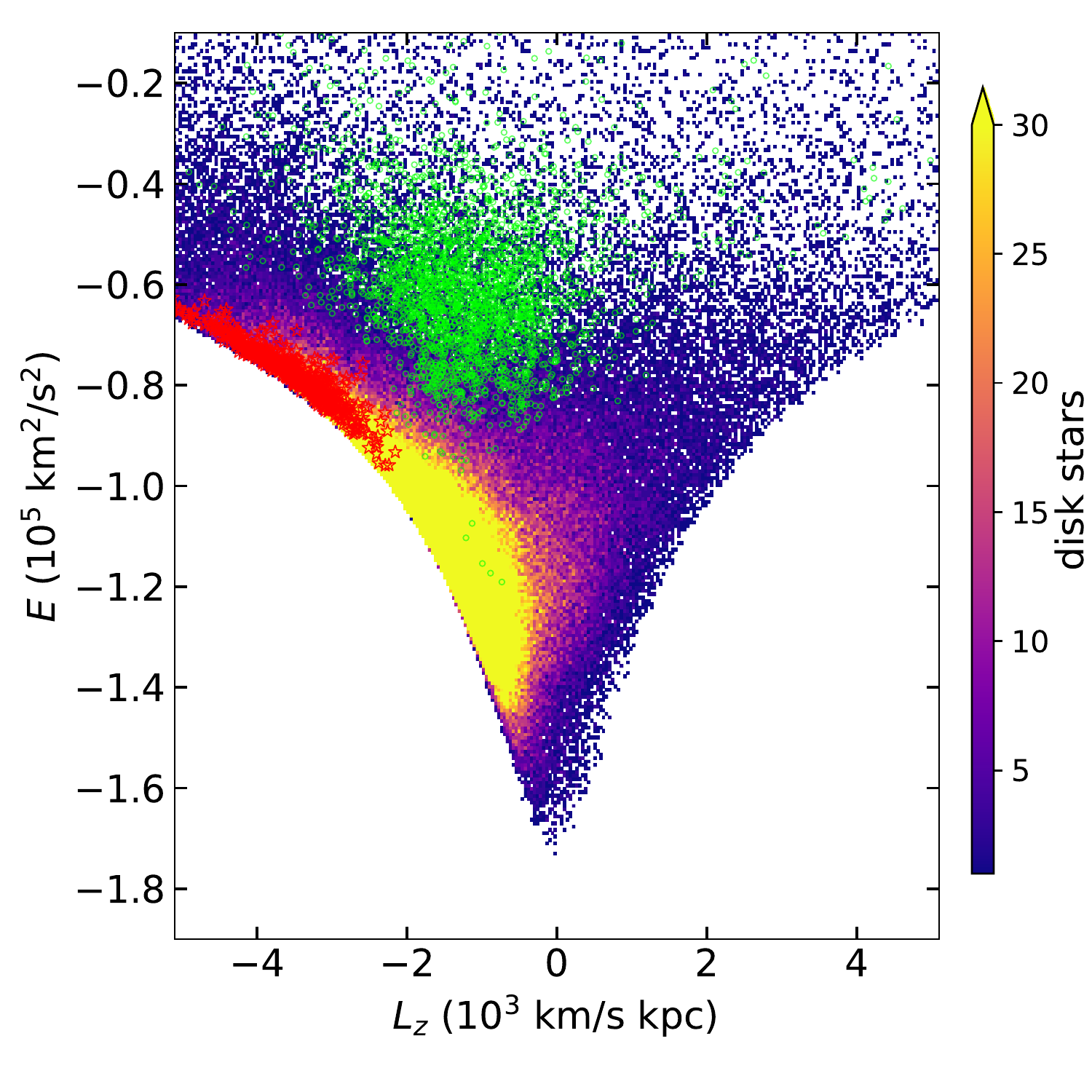}
\caption{E-L$_{Z}$ density distribution for Galactic disc stars (coloured density, the data is from the K giants catalogue with $|Z|<3$ kpc.), GASS members (red stars), and Sagittarius stream members selected from \citet{2019ApJ...886..154Y} (green circles).}
\label{LE}
\end{figure}

\begin{figure}
\centering
\includegraphics[width=0.5\columnwidth]{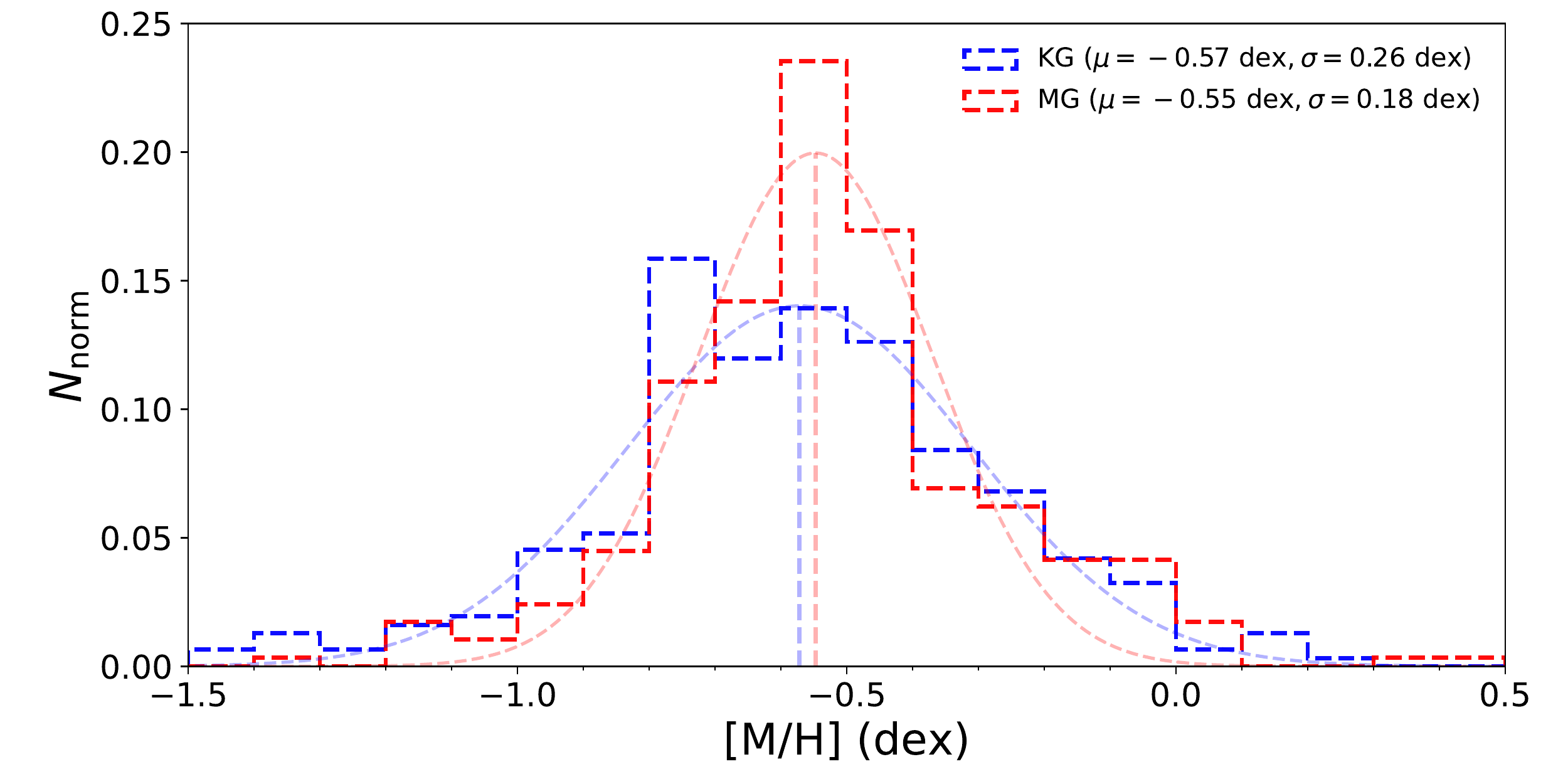}
\includegraphics[width=0.5\columnwidth]{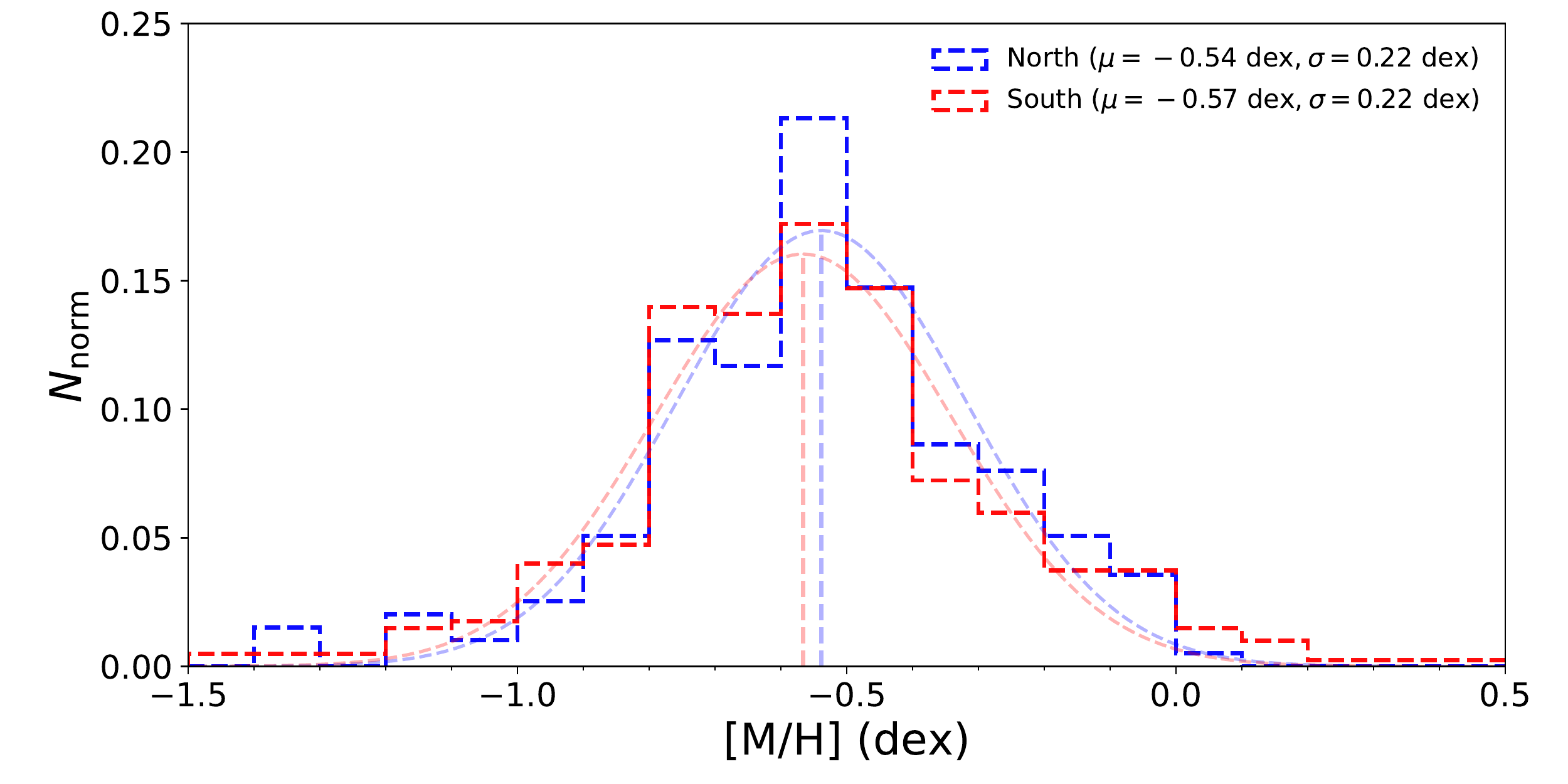}
\caption{Metallicity distribution for our GASS members. The left panel shows a comparison between the M giants (red line) and the K giants (blue dash line). The right panel shows comparison between the south and north samples.}
\label{mh}
\end{figure}

\begin{figure}
\includegraphics[width=\columnwidth]{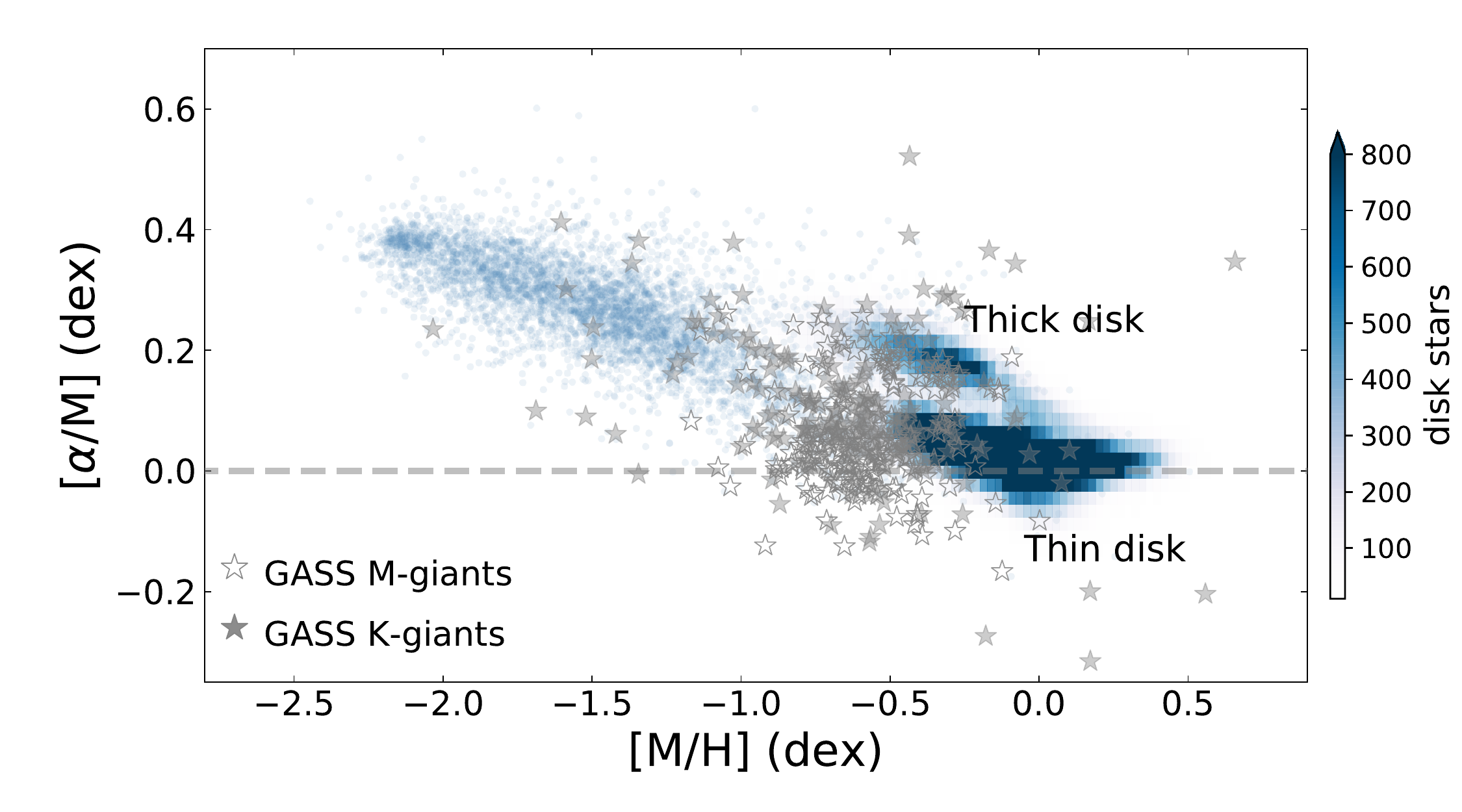}
\caption{[$\alpha$/M] versus [M/H] distribution of stars from GASS (grey stars), compared to the thick disk, thin disk, and halo stars (blue density). 
}
\label{alpha}
\end{figure}

\begin{figure}
\centering
\includegraphics[width=0.9\columnwidth]{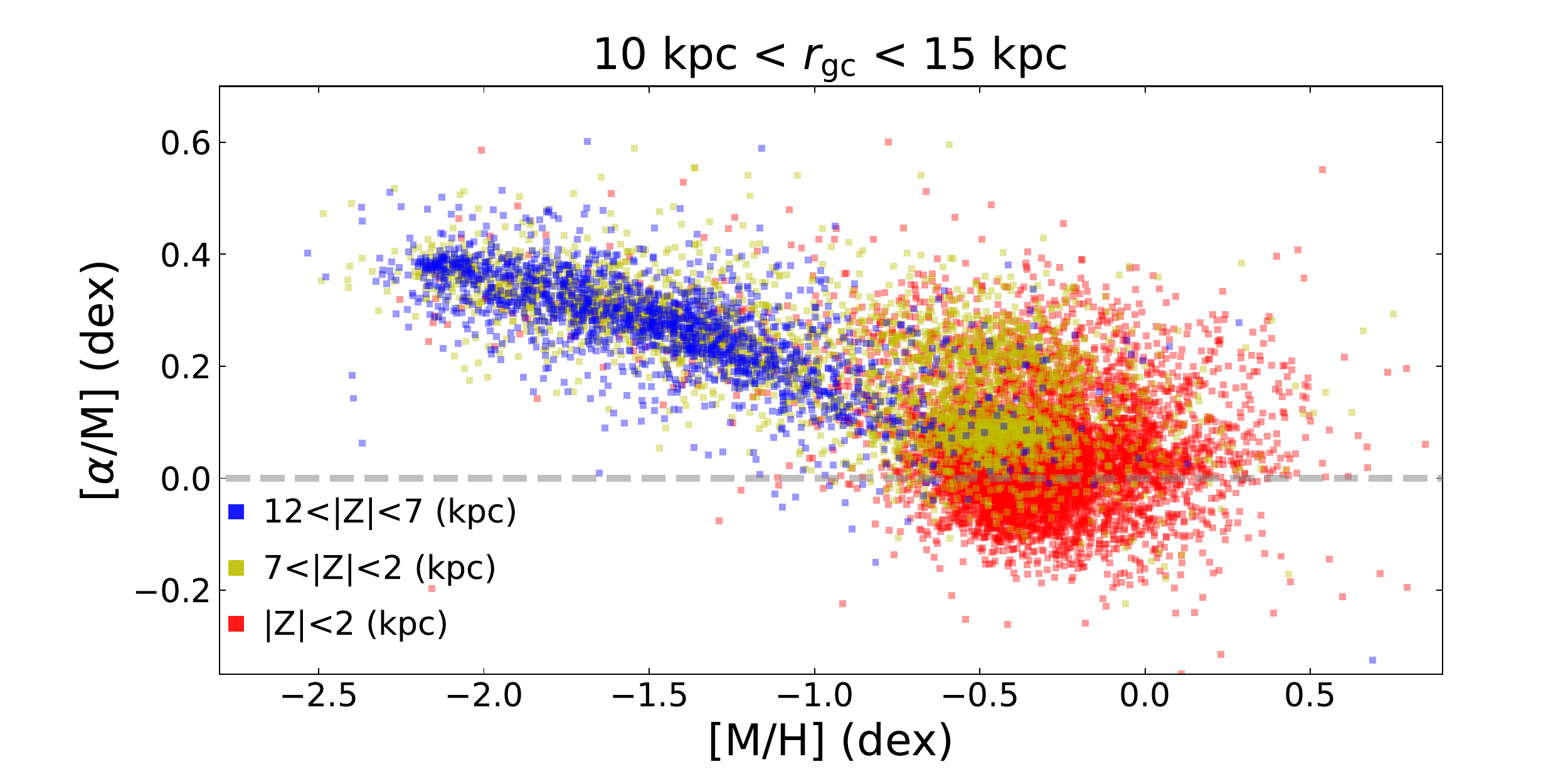}
\includegraphics[width=0.9\columnwidth]{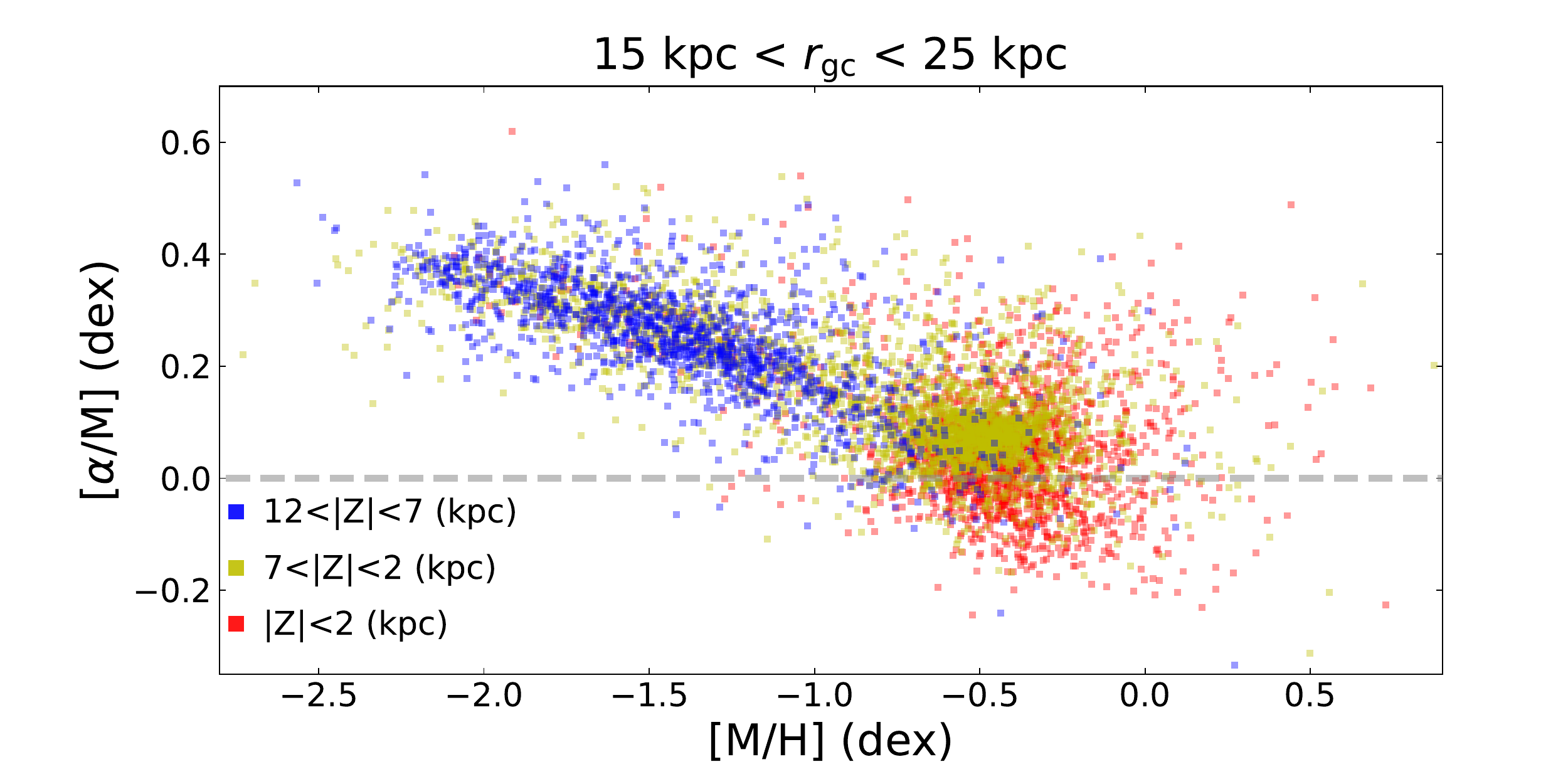} 
\caption{[$\alpha$/M] versus [M/H] distribution of the K-giants sample. The upper panel shows all stars selected between $10 < r_{GC} < 15$ kpc but where we have subtracted the group members identified in XXXue19, and then split into samples of different distance to the Galactic plane: $7<Z<12$ kpc, $2 <Z< 7$ kpc, $-7 < Z < -2$ kpc, $-12 < Z < -7$ kpc, and a ''disk'' sample at $-2 <Z< 2$ kpc. The lower panel is similar to the upper panel but for a different distance range of $15 < r_{GC} < 25$ kpc.}
\label{compare_sample}
\end{figure}

 The velocity vector directions of GASS in the X-Y and Y-Z planes indicate that the GASS consists of two circle orbits on both sides of the MW disk which span a large distances from 15 kpc to 30 kpc.{\textbf{We can see clear velocity gradient on the Y-Z plane as shown in Figure~\ref{yz}, }}with the $V_Z$ pointing toward and outward from the mid-plane at different distances in the southern hemisphere. 
 We also compared our GASS stars to the disk and Sagittarius stream in $E$-$L_{z}$ space. GASS members have a similar L-E distribution to the thin disk distribution. 
 
 We also present the metallicity distribution of GASS. The total [M/H] distributions for K- and M-giant GASS members are similar. We did not find significant [M/H] differences for the GASS south and north rings. By comparing the $\alpha$-abundance with the Galactic components, the trend of [$\alpha$/M] is neither the same as the traditional Galactic disk nor halo populations.  The GASS distribution in [$\alpha$/M] vs [M/H] space is consistent with the continuation of metal-rich thin disk stars into the outer halo \citep{2016A&A...589A..66H}. It is also consistent with the $\alpha$-abundance derived in \citet{2018ApJ...859L...8H} for the TriAnd substructure, which suggests that this feature is an "extension" of the trend seen in the disk.
 
Our analysis shows that Mon, TriAnd, and A13 (and possibly including other stars near the Galactic anticenter) have similar kinematic and chemical features. These stars may be just part of the outer-disk, with this outer disk extending out to at least $r_{gc}\sim$30 kpc.  GASS may have formed in the outer disk where there may still be more cold gas in the present day and then maybe lower star formation efficiency than the inner disk. 
{\textbf{We also can infer that the GASS stars may formed after the thick disk was formed because the molecular cloud density decreased in the outer disk where the SFR might be less efficient than the inner disk.}}

\acknowledgments
This work is supported by NSFC grant No. 11988101, 11873052, 11890694, 11835057, 11703019,11503066, U1731129 and 11703038, by the National Key R\&D Program of China under grant No. 2019YFA0405500, and by China West Normal University grants 17C053, 17YC507 and 16EE018. JLC acknowledges support from HST grant HST-GO-15228.001-A and NSF grant AST-1816196. R.A.M. acknowledges support from the Chilean Centro de Excelencia en Astrofisica y Tecnologias Afines (CATA) BASAL AFB-170002, and FONDECYT/CONICYT grant \# 1190038. This work is sponsored (in part) by the Chinese Academy of Sciences (CAS), through a grant to the CAS South America Center for Astronomy (CASSACA) in Santiago, Chile. 
 
Guoshoujing Telescope (the Large Sky Area Multi-Object Fiber Spectroscopic Telescope LAMOST) is a National Major Scientific Project built by the Chinese Academy of Sciences. Funding for the project has been provided by the National Development and Reform Commission. LAMOST is operated and managed by the National Astronomical Observatories, Chinese Academy of Sciences. 
 
This work has made use of data from the European Space Agency (ESA) mission {\it Gaia} (\url{https://www.cosmos.esa.int/gaia}), processed by the {\it Gaia} Data Processing and Analysis Consortium (DPAC, \url{https://www.cosmos.esa.int/web/gaia/dpac/consortium}). Funding
for the DPAC has been provided by national institutions, in particular the institutions participating in the {\it Gaia} Multilateral Agreement.

\begin{sidewaystable}[htb]
\caption{Parameters of GASS Stars} \label{t_catalog}
\footnotesize
\setlength\tabcolsep{4pt}
\begin{tabular}{@{\extracolsep{-0.1cm}}cclrr ccr@{\hspace{0.3cm}}r@{\hspace{0.5cm}}c ccccc cc}
\tablewidth{0pt}
\hline
\hline
\colhead{LAMOST\tablenotemark{a}}         &
\colhead{Gaia\tablenotemark{b}}                &
\colhead{Type}                &
\colhead{R.A.}                &
\colhead{Decl.}               &
\colhead{$d$}                 &
\colhead{$\Delta d$}              &
\colhead{$hrv$}               &
\colhead{$\Delta hrv$}            &
\colhead{pmra}                &
\colhead{$\Delta$pmra}             &
\colhead{pmdec}               &
\colhead{$\Delta$pmdec}            &
\colhead{[M/H]}              &
\colhead{$\Delta$[M/H]}           &
\colhead{[$\alpha$/M]}       &
\colhead{$\Delta [\alpha$/M]}
\\
\colhead{} &
\colhead{} &
\colhead{} &
\colhead{deg} &
\colhead{deg} &
\colhead{kpc} &
\colhead{kpc} &
\colhead{km s$^{-1}$} &
\colhead{km s$^{-1}$} &
\colhead{mas yr$^{-1}$} &
\colhead{mas yr$^{-1}$} &
\colhead{mas yr$^{-1}$} &
\colhead{mas yr$^{-1}$} &
\colhead{dex} &
\colhead{dex} &
\colhead{dex} &
\colhead{dex}
\\
\hline
216068    & 352530915661312     & LAMOST KG & 44.017706  & 0.962512  & 15.6 & 1.6 & $  1.7$  &  5.1 & $0.142$  & 0.082 & $-0.967$ & 0.076 & $-0.21$ & 0.11 & $ 0.04$ & 0.05  \\
715099    & 2872282645040005120 & LAMOST KG & 352.857183 & 32.318859 & 19.2 & 2.8 & $-171.9$ &  5.5 & $ -0.72$ & 0.082 & $-0.929$ & 0.059 & $-0.79$ & 0.23 & $ 0.07$ & 0.07  \\
716170    & 2872538315852901632 & LAMOST KG & 352.093575 & 32.856964 & 15.3 & 2.5 & $-145.5$ &  6.7 & $-1.055$ & 0.074 & $-0.926$ & 0.056 & $-0.40$ & 0.23 & $-0.07$ & 0.07  \\
1107119   & 115451783231530112  & LAMOST KG & 47.05329   & 26.617279 & 14.6 & 2.1 & $ -44.4$ &  6.8 & $0.205$  & 0.204 & $-0.571$ & 0.141 & $-0.77$ & 0.22 & $ 0.11$ & 0.07  \\
1202158   & 1888486850788253056 & LAMOST KG & 337.893154 & 29.742142 & 20.6 & 1.3 & $-199.0$ &  4.7 & $-1.203$ & 0.077 & $-1.081$ & 0.084 & $-1.0$  & 0.19 & $ 0.04$ & 0.07  \\
1216219   & 1901404634945395456 & LAMOST KG & 338.411149 & 32.640892 & 16.9 & 4.4 & $-191.6$ &  4.5 & $-1.253$ & 0.09  & $-1.193$ & 0.081 & $-0.78$ & 0.16 & $ 0.07$ & 0.06  \\
1315042   & 2867487021995351040 & LAMOST KG & 357.426356 & 30.115    & 16.3 & 2.1 & $-197.8$ &  3.6 & $-0.506$ & 0.037 & $-1.074$ & 0.027 & $-0.70$ & 0.13 & $-0.09$ & 0.05  \\
1315180   & 2867075525473014528 & LAMOST KG & 357.18861  & 41.642576 & 19.0 & 2.1 & $-143.9$ &  7.3 & $-0.949$ & 0.088 & $-0.944$ & 0.057 & $-0.39$ & 0.25 & $ 0.08$ & 0.08  \\
1609069   & 375247964054278912  & LAMOST KG & 12.316544  & 29.804014 & 14.4 & 4.0 & $-155.4$ &  4.3 & $-0.602$ & 0.077 & $-0.403$ & 0.069 & $-0.47$ & 0.27 & $ 0.09$ & 0.08  \\
7704197   & 311445949992518912  & LAMOST KG & 15.543772  & 30.905615 & 12.2 & 3.1 & $-117.3$ &  5.2 & $-0.794$ & 0.057 & $-0.716$ & 0.052 & $-0.60$ & 0.17 & $ 0.11$ & 0.06  \\
\hline
\multicolumn{4}{l}{\textsuperscript{\hspace{-0.5em} a} Unique identifier in LAMOST.}\\
\multicolumn{4}{l}{\textsuperscript{\hspace{-0.5em} b} Solution identifier in Gaia.}\\
\multicolumn{4}{l}{(This table is available in its entirety in machine-readable form.)}
\end{tabular}
\end{sidewaystable}

\begin{sidewaystable}[htb]
\caption{Orbital Parameters of GASS Stars} \label{t_orbs}
\scriptsize 
\setlength\tabcolsep{10pt}
\begin{tabular}{ccccc ccrrc ccccc}
\tablewidth{0pt}
\hline
\hline
\colhead{LAMOST} &
\colhead{$e$} &
\colhead{$\Delta e$} &
\colhead{$a$} &
\colhead{$\Delta a$} &
\colhead{$l_{\rm{orb}}$} &
\colhead{$\Delta l_{\rm{orb}}$} &
\colhead{$b_{\rm{orb}}$} &
\colhead{$\Delta b_{\rm{orb}}$} &
\colhead{$l_{\rm{apo}}$} &
\colhead{$\Delta l_{\rm{apo}}$} &
\colhead{$E$} &
\colhead{$\Delta E$} &
\colhead{$L$}&
\colhead{$\Delta L$}
\\
\colhead{ } &
\colhead{ } &
\colhead{ } &
\colhead{kpc} &
\colhead{kpc} &
\colhead{deg} &
\colhead{deg} &
\colhead{deg} &
\colhead{deg} &
\colhead{deg} &
\colhead{deg} &
\colhead{km$^2$ s$^{-2}$} &
\colhead{km$^2$ s$^{-2}$} &
\colhead{km s$^{-1}$ kpc} &
\colhead{km s$^{-1}$ kpc}
\\
\hline
216068    & 19.20 & 1.21 & 0.13 & 0.04 & 336.18 &  4.05 & 145.34 & 1.68 & 318.24 & 12.85 & $-78274.00$ & 1975.62 & 3696.46 & 198.19 \\
715099    & 19.10 & 3.63 & 0.23 & 0.06 & 326.94 & 13.01 & 155.08 & 0.77 & 322.60 & 73.65 & $-78460.68$ & 5729.74 & 3555.62 & 661.74 \\
716170    & 16.91 & 3.66 & 0.13 & 0.06 & 338.98 & 32.60 & 155.52 & 0.66 & 348.73 &145.56 & $-82678.12$ & 6309.37 & 3294.14 & 614.95 \\
1107119   & 21.23 & 3.42 & 0.11 & 0.06 & 351.15 &148.91 & 162.12 & 1.77 & 355.07 &143.03 & $-76778.87$ & 4923.66 & 3892.91 & 544.48 \\
1202158   & 20.94 & 4.51 & 0.24 & 0.04 & 331.49 &  9.03 & 152.31 & 1.92 & 345.07 & 85.01 & $-75428.06$ & 6415.99 & 3846.37 & 748.22 \\
1216219   & 16.16 & 3.19 & 0.28 & 0.06 & 336.28 & 23.05 & 156.04 & 1.71 & 336.99 &137.39 & $-83908.12$ & 6040.82 & 3019.25 & 588.88 \\
1315042   & 15.19 & 1.16 & 0.45 & 0.03 & 335.95 &  8.67 & 153.21 & 0.59 & 269.74 & 64.04 & $-85439.85$ & 2511.87 & 2571.41 & 218.39 \\
1315180   & 23.58 & 9.37 & 0.13 & 0.07 & 324.86 &  20.8 & 152.86 & 0.79 & 354.57 & 68.22 & $-72894.93$ & 9918.26 & 4300.86 & 1264.50 \\
1609069   & 17.83 & 2.54 & 0.20 & 0.04 & 355.38 &157.38 & 162.05 & 0.81 & 320.28 &161.97 & $-80862.51$ & 4497.49 & 3383.99 & 453.92 \\
7704197   & 17.23 & 2.73 & 0.13 & 0.03 & 347.19 & 81.99 & 156.75 & 1.01 & 359.75 & 56.65 & $-81949.20$ & 4950.32 & 3360.89 & 458.79 \\
\hline
\multicolumn{8}{l}{(This table is available in its entirety in machine-readable form.)}
\end{tabular}
\end{sidewaystable}




\begin{thebibliography}{}
\bibitem[An et al.(2013)]{2013ApJ...763...65A} An, D., Beers, T.~C., Johnson, J.~A., et al.\ 2013, \apj, 763, 65. doi:10.1088/0004-637X/763/1/65
\bibitem[Antoja et al.(2018)]{2018Natur.561..360A} Antoja, T., Helmi, A., Romero-G{\'o}mez, M., et al.\ 2018, \nat, 561, 360
\bibitem[Arifyanto et al.(2005)]{2005A&A...433..911A} Arifyanto, M.~I., Fuchs, B., Jahrei{\ss}, H., et al.\ 2005, \aap, 433, 911. doi:10.1051/0004-6361:20035829

\bibitem[Bailer-Jones et al.(2018)]{2018AJ....156...58B} Bailer-Jones, C.~A.~L., Rybizki, J., Fouesneau, M., Mantelet, G., \& Andrae, R.\ 2018, \aj, 156, 58 
\bibitem[Beers \& Sommer-Larsen(1995)]{1995ApJS...96..175B} Beers, T.~C. \& Sommer-Larsen, J.\ 1995, \apjs, 96, 175. doi:10.1086/192117
\bibitem[Beers et al.(2012)]{2012ApJ...746...34B} Beers, T.~C., Carollo, D., Ivezi{\'c}, {\v{Z}}., et al.\ 2012, \apj, 746, 34. doi:10.1088/0004-637X/746/1/34
\bibitem[Beers et al.(2000)]{2000AJ....119.2866B} Beers, T.~C., Chiba, M., Yoshii, Y., et al.\ 2000, \aj, 119, 2866. doi:10.1086/301410
\bibitem[Beers et al.(1995)]{1995ApJs...96...175} Beers, T.~C., Norris, \& Sommer-Larsen,J.\ 1995, \apjs, 96, 175
\bibitem[Beers et al.(2014)]{2014ApJ...794...58B} Beers, T.~C., Norris, J.~E., Placco, V.~M., et al.\ 2014, \apj, 794, 58
\bibitem[Bensby et al.(2003)]{2003A&A...410..527B} Bensby, T., Feltzing, S., \& Lundstr{\"o}m, I.\ 2003, \aap, 410, 527. doi:10.1051/0004-6361:20031213
\bibitem[Bensby et al.(2014)]{2014A&A...562A..71B} Bensby, T., Feltzing, S., \& Oey, M.~S.\ 2014, \aap, 562, A71. doi:10.1051/0004-6361/201322631
\bibitem[Bidelman \& MacConnell(1973)]{1973AJ.....78..687B} Bidelman, W.~P. \& MacConnell, D.~J.\ 1973, \aj, 78, 687. doi:10.1086/111475
\bibitem[Binney \& Sch{\"o}nrich(2018)]{2018MNRAS.481.1501B} Binney, J., \& Sch{\"o}nrich, R.\ 2018, \mnras, 481, 1501 
\bibitem[Bird et al.(2019)]{Bird2019} Bird, S.~A., Xue, X.-X., Liu, C., et al.\ 2019, \aj, 157, 104
\bibitem[Bland-Hawthorn et al.(2019)]{2019MNRAS.486.1167B} Bland-Hawthorn, J., Sharma, S., Tepper-Garcia, T., et al.\ 2019, \mnras, 486, 1167 

\bibitem[Carlin et al.(2014)]{2014AAS...22334613C} Carlin, J.~L., DeLaunay, J., Newberg, H.~J., et al.\ 2014, American Astronomical Society Meeting Abstracts \#223
\bibitem[Carraro \& Costa(2009)]{2009A&A...493...71C} Carraro, G., \& Costa, E.\ 2009, \aap, 493, 71 
\bibitem[Carollo et al.(2007)]{2007Natur.450.1020C} Carollo, D., Beers, T.~C., Lee, Y.~S., et al.\ 2007, \nat, 450, 1020. doi:10.1038/nature06460
\bibitem[Carollo et al.(2010)]{2010ApJ...712..692C} Carollo, D., Beers, T.~C., Chiba, M., et al.\ 2010, \apj, 712, 692. doi:10.1088/0004-637X/712/1/692

\bibitem[Carollo et al.(2012)]{2012ApJ...744..195C} Carollo, D., Beers, T.~C., Bovy, J., et al.\ 2012, \apj, 744, 195. doi:10.1088/0004-637X/744/2/195
\bibitem[Carollo et al.(2014)]{2014ApJ...788..180C} Carollo, D., Freeman, K., Beers, T.~C., et al.\ 2014, \apj, 788, 180. doi:10.1088/0004-637X/788/2/180
\bibitem[Cheng et al.(2019)]{2019ApJ...872L...1C} Cheng, X., Liu, C., Mao, S., et al.\ 2019, \apjl, 872, L1
\bibitem[Cheng et al.(2012)]{2012ApJ...752...51C} Cheng, J.~Y., Rockosi, C.~M., Morrison, H.~L., et al.\ 2012, \apj, 752, 51. doi:10.1088/0004-637X/752/1/51
\bibitem[Chiba \& Yoshii(1998)]{1998AJ....115..168C} Chiba, M. \& Yoshii, Y.\ 1998, \aj, 115, 168. doi:10.1086/300177


\bibitem[Chou et al.(2010)]{2010ApJ...720L...5C} Chou, M.-Y., Majewski, S.~R., Cunha, K., et al.\ 2010, \apjl, 720, L5 
\bibitem[Cui et al. (2012)]{Cui2012} Cui, X.-Q., Zhao, Y.-H., Chu, Y.-Q., et al., RAA, 12, 1197
\bibitem[Deason et al.(2014)]{2014MNRAS.444.3975D} Deason, A.~J., Belokurov, V., Hamren, K.~M., et al.\ 2014, \mnras, 444, 3975 
\bibitem[de Grijs \& Bono(2016)]{2016ApJS..227....5D} de Grijs, R., \& Bono, G.\ 2016, \apjs, 227, 5 
\bibitem[de Grijs \& Bono(2017)]{2017ApJS..232...22D} de Grijs, R., \& Bono, G.\ 2017, \apjs, 232, 22 
\bibitem[Deng et al. (2012)]{Deng2012} Deng, L.-C., Newberg, H.J., Liu, C., et al., 2012, RAA, 12, 735
\bibitem[Feast et al.(2014)]{2014Natur.509..342F} Feast, M.~W., Menzies, J.~W., Matsunaga, N., \& Whitelock, P.~A.\ 2014, \nat, 509, 342 
\bibitem[Gaia Collaboration et al.(2018)]{gaia2018}Gaia Collaboration, Brown, A. G. A., Vallenari, A., et al.\ 2018, ArXiv e-prints, arXiv:1804.09365
\bibitem[Gilmore et al.(2002)]{2002ApJ...574L..39G} Gilmore, G., Wyse, R.~F.~G., \& Norris, J.~E.\ 2002, \apjl, 574, L39. doi:10.1086/342363
\bibitem[G{\'o}mez et al.(2016)]{gomez2016} G{\'o}mez, F.~A., White, S.~D.~M., Marinacci, F., et al.\ 2016, \mnras, 456, 2779 
\bibitem[Hammersley \& L{\'o}pez-Corredoira(2011)]{2011A&A...527A...6H} Hammersley, P.~L., \& L{\'o}pez-Corredoira, M.\ 2011, \aap, 527, A6 
\bibitem[Hayes et al.(2018)]{2018ApJ...859L...8H} Hayes, C.~R., Majewski, S.~R., Hasselquist, S., et al.\ 2018, \apjl, 859, L8 
\bibitem[Haywood et al.(2016)]{2016A&A...589A..66H} Haywood, M., Lehnert, M.~D., Di Matteo, P., et al.\ 2016, \aap, 589, A66
\bibitem[Hernquist(1990)]{1990ApJ...356..359H} Hernquist, L.\ 1990, \apj, 356, 359 
\bibitem[Ibata et al.(2003)]{2003MNRAS.340L..21I} Ibata, R.~A., Irwin, M.~J., Lewis, G.~F., Ferguson, A.~M.~N., \& Tanvir, N.\ 2003, \mnras, 340, L21
\bibitem[Ji et al.(2016)]{ji2016} Ji, W., Cui, W., Liu, C., et al.\ 2016, \apjs, 226, 1 
\bibitem[Laporte et al.(2018)]{laporte2018} Laporte, C.~F.~P., G{\'o}mez, F.~A., Besla, G., et al.\ 2018, \mnras, 473, 1218
\bibitem[Laporte et al.(2019)]{2019MNRAS.485.3134L} Laporte, C.~F.~P., Minchev, I., Johnston, K.~V., \& G{\'o}mez, F.~A.\ 2019, \mnras, 485, 3134 
\bibitem[Li et al.(2012)]{2012ApJ...757..151L} Li, J., Newberg, H.~J., Carlin, J.~L., et al.\ 2012, \apj, 757, 151 
\bibitem[Li et al.(2016)]{li2016} Li, J., Smith, M.~C., Zhong, J., et al.\ 2016, \apj, 823, 59 
\bibitem[Li et al.(2017)]{liting2017} Li, T.~S., Sheffield, A.~A., Johnston, K.~V., et al.\ 2017, \apj, 844, 74 
\bibitem[Li et al.(2018)]{lyb2018} Li, Y.-B., Luo, A.-L., Du, C.-D., et al.\ 2018, \apjs, 234, 31 
\bibitem[Li \& Zhao(2017)]{2017ApJ...850...25L} Li, C. \& Zhao, G.\ 2017, \apj, 850, 25
\bibitem[Liu et al.(2014)]{liu2014} Liu, C., Deng, L.-C., Carlin, J.~L., et al.\ 2014, \apj, 790, 110
\bibitem[Liu et al.(2017)]{2017ApJ...835L..18L} Liu, C., Wang, Y.-G., Shen, J., et al.\ 2017, \apjl, 835, L18
\bibitem[L{\'o}pez-Corredoira et al.(2002)]{2002A&A...394..883L} L{\'o}pez-Corredoira, M., Cabrera-Lavers, A., Garz{\'o}n, F., \& Hammersley, P.~L.\ 2002, \aap, 394, 883 
\bibitem[Luo et al.(2012)]{Luo2012} Luo, A.-L., Zhang, H.-T., Zhao, Y.-H., et al.\ 2012, Research in Astronomy and Astrophysics, 12, 1243 
\bibitem[Majewski et al.(2017)]{Majewski17} Majewski, S.~R., Schiavon, R.~P., Frinchaboy, P.~M., et al.\ 2017, \aj, 154, 94 
\bibitem[Martin \& Morrison(1998)]{1998AJ....116.1724M} Martin, J.~C. \& Morrison, H.~L.\ 1998, \aj, 116, 1724. doi:10.1086/300568
\bibitem[McConnachie(2012)]{2012AJ....144....4M} McConnachie, A.~W.\ 2012, \aj, 144, 4. doi:10.1088/0004-6256/144/1/4
\bibitem[Meisner et al.(2012)]{2012ApJ...753..116M} Meisner, A.~M., Frebel, A., Juri{\'c}, M., \& Finkbeiner, D.~P.\ 2012, \apj, 753, 116 
\bibitem[Moitinho et al.(2006)]{2006MNRAS.368L..77M} Moitinho, A., V{\'a}zquez, R.~A., Carraro, G., et al.\ 2006, \mnras, 368, L77 
\bibitem[Momany et al.(2006)]{2006A&A...451..515M} Momany, Y., Zaggia, S., Gilmore, G., et al.\ 2006, \aap, 451, 515 
\bibitem[Morrison et al.(1990)]{1990AJ....100.1191M} Morrison, H.~L., Flynn, C., \& Freeman, K.~C.\ 1990, \aj, 100, 1191
\bibitem[Morrison et al.(2009)]{2009ApJ...694..130M} Morrison, H.~L., Helmi, A., Sun, J., et al.\ 2009, \apj, 694, 130. doi:10.1088/0004-637X/694/1/130
\bibitem[Navarro et al.(1996)]{Navarro1996} Navarro, J.~F., Frenk, C.~S., \& White, S.~D.~M.\ 1996, \apj, 462, 563 
\bibitem[Newberg \& Carlin(2016)]{2016N} Newberg, H.~J., \& Carlin, J.~L.\ 2016, Astrophysics and Space Science Library, 420,  
\bibitem[Newberg et al.(2002)]{nyetal02} Newberg, H.~J., Yanny, B., Rockosi, C., et al.\ 2002, \apj, 569, 245
\bibitem[Norris et al.(1985)]{1985ApJS...58..463N} Norris, J., Bessell, M.~S., \& Pickles, A.~J.\ 1985, \apjs, 58, 463
\bibitem[Rocha-Pinto et al.(2003)]{RP2003} Rocha-Pinto, H.~J., Majewski, S.~R., Skrutskie, M.~F., \& Crane, J.~D.\ 2003, \apjl, 594, L115 
\bibitem[Sch{\"o}nrich et al.(2010)]{2010MNRAS.403.1829S} Sch{\"o}nrich, R., Binney, J., \& Dehnen, W.\ 2010, \mnras, 403, 1829 
\bibitem[Sharma et al.(2010)]{2010ApJ...722..750S} Sharma, S., Johnston, K.~V., Majewski, S.~R., et al.\ 2010, \apj, 722, 750 
\bibitem[Sheffield et al.(2018)]{2018ApJ...854...47S} Sheffield, A.~A., Price-Whelan, A.~M., Tzanidakis, A., et al.\ 2018, \apj, 854, 47 
\bibitem[Simon(2019)]{2019ARA&A..57..375S} Simon, J.~D.\ 2019, \araa, 57, 375. doi:10.1146/annurev-astro-091918-104453
\bibitem[Slater et al.(2014)]{2014ApJ...791....9S} Slater, C.~T., Bell, E.~F., Schlafly, E.~F., et al.\ 2014, \apj, 791, 9
\bibitem[Steinmetz et al.(2006)]{2006AJ....132.1645S} Steinmetz, M., Zwitter, T., Siebert, A., et al.\ 2006, \aj, 132, 1645. doi:10.1086/506564
\bibitem[Tian et al.(2018)]{2018ApJ...865L..19T} Tian, H.-J., Liu, C., Wu, Y., et al.\ 2018, \apjl, 865, L1
\bibitem[Villalobos \& Helmi(2009)]{2009MNRAS.399..166V} Villalobos, {\'A}. \& Helmi, A.\ 2009, \mnras, 399, 166. doi:10.1111/j.1365-2966.2009.15085.x
\bibitem[Wang et al.(2018)]{2018MNRAS.478.3367W} Wang, H.-F., Liu, C., Xu, Y., et al.\ 2018, \mnras, 478, 3367. doi:10.1093/mnras/sty1058
\bibitem[Wang et al.(2020)]{2020MNRAS.491.2104W} Wang, H.-F., L{\'o}pez-Corredoira, M., Huang, Y., et al.\ 2020, \mnras, 491, 2104. doi:10.1093/mnras/stz3113
\bibitem[Xu et al.(2015)]{2015ApJ...801..105X} Xu, Y., Newberg, H.~J., Carlin, J.~L., et al.\ 2015, \apj, 801, 105 
\bibitem[Xue et al.(2014)]{Xue14} Xue, X.-X., Ma, Z., Rix, H.-W., et al.\ 2014, \apj, 784, 170 
\bibitem[Yang et al.(2019)]{2019ApJ...886..154Y} Yang, C., Xue, X.-X., Li, J., et al.\ 2019, \apj, 886, 154
\bibitem[Yanny et al.(2003)]{2003Yanny} Yanny, B., Newberg, H.~J., Grebel, E.~K., et al.\ 2003, \apj, 588, 824 
\bibitem[Yanny \& Newberg(2016)]{YannyNewberg2016} Yanny, B., \& Newberg, H.~J.\ 2016, Tidal Streams in the Local Group and Beyond, 63
\bibitem[Yang et al.(2019)]{Yang2019b} Yang, C., Xue, X.-X., Li, J., et al.\ 2019, \apj, 886, 154
\bibitem[Yang et al.(2019)]{Yang2019a} Yang, C., Xue, X.-X., Li, J., et al.\ 2019, \apj, 880, 65
\bibitem[York et al.(2000)]{2000AJ....120.1579Y} York, D.~G., Adelman, J., Anderson, J.~E., et al.\ 2000, \aj, 120, 1579. doi:10.1086/301513
\bibitem[Zhao et al.(2012)]{Zhao2012} Zhao, G., Zhao, Y.-H., Chu,
  Y.-Q., et al. \ 2012, RAA, 12, 723
\bibitem[Zhang et al.(2019)]{Zhang2019} Zhang, B., Liu, C., Deng,Licai. \ 2019, arXiv:1908.08677
\bibitem[Zhong \& Li et al.(2019)]{Zhong2019b} Zhong, J., Li, J., Carlin, L. Jeffery., et al. \ 2019,\apjs,244,8









\end{thebibliography}
\end{document}